\documentclass[a4 paper, superscriptaddress, 12pt]{article}

\usepackage[utf8]{inputenc}
\usepackage[T1]{fontenc}
\usepackage{amsmath, amsfonts, amssymb}
\usepackage{graphicx}
\usepackage{fancyhdr}
\usepackage[left=3cm, right=3cm, top=3cm, bottom=3cm]{geometry}
\usepackage{enumitem}
\usepackage{multirow}
\usepackage[numbers,sort&compress]{natbib}% Citation support using natbib.sty
\usepackage{braket}

\newcommand{\La}{\mathcal{L}}

\begin{document}
\begin{center}
\textbf{\Large{The Higgs Boson and The Fakeon Hypothesis}}
\end{center}
%
%%%
%
\begin{center}
\textrm{\normalsize{\textbf{$^\ast$Musongela\  Lubo, $^{\dagger\S}$Muanza\ Steve, $^\ast$Kikunga\  Kasenda\ Ivan, $^\ast$Mavungu\  Tsava\ Christian}}}\\
\end{center}
\begin{center}
{\textit{\scriptsize{$^\star$Physics Department, Faculty of Sciences, University of Kinshasa, P.O. Box 190 Kin\ XI, Kinshasa, D.R.Congo}}}\\
{\textit{\scriptsize{$^\dagger$General Commission for Atomic Energy, Regional Center for Nuclear Studies in Kinshasa,}}} {\textit{\scriptsize{Campus of the University of Kinshasa, P.O. Box 868 Kin\ XI, Kinshasa, D.R.Congo}}}\\
{\textit{\scriptsize{$^{\S}$Aix-Marseille University, 163 Avenue de Luminy, 13288 Marseille Cedex 09, France}}}
\end{center}
\begin{center}
{\scriptsize{Email: musongela.lubo@unikin.ac.cd, muanza@cern.ch, ivan.kikunga@unikin.ac.cd, Christiantsava333@gmail.com}}\\
\end{center}
%
%%%
%
\begin{abstract}
	In this paper, we make an attempt to implement the fakeon hypothesis in particle physics. To begin with, we consider a model in which the Higgs boson is the only fakeon. We deduce its interactions with the electroweak gauge bosons. Each such new interaction can be written as a product of two factors. The first one depends, on the electroweak gauge bosons and their derivatives. The second one solely depends, on the physical Higgs and its derivatives. We also study the conserved quantities of different (free) fields in this setting.
\end{abstract}


%%%%%%%%%%%%%%%%%%%%%%%%%%%%%%%%%%%%%%%%%%%%%%%%%%%%%%%%%%%%%%%%%%%%%%%%%%%%%%%%%%%%%%%%%%%%%%%%%%%%%%%%%%%%%%%%%%%%%%%%%%%%%
%%%%%%%%%%%%%%%%%%%%%%%%%%%%%%%%%%%%%%%%%%%%%%%%%%%%%%%%%%%%%%%%%%%%%%%%%%%%%%%%%%%%%%%%%%%%%%%%%%%%%%%%%%%%%%%%%%%%%%%%%%%%%

\section{Introduction}
\label{sec01}
\

The Standard Model of particle physics is one of the most successful constructions of theoretical physics. However, no knowledge is definitive. Thus, different works go beyond that model. Of course, the results must be in accordance with the well verified experimental results described by that model.\\

Studies beyond the Standard Model are of many kinds. This model is basically built upon two bases : the gauge group $SU(3)_{C} \times SU(2)_{L} \times U(1)_{Y}$ and the Poincar\'e group. The first group is about internal symmetries while the second one concerns space-time transformations. Considerations beyond the Standard Model  concerns mostly the change of one of these ingredients, and possibly the particle content.\\

One of the possibilities is to keep the Lorentz group but to change the internal group. The first attempt is to replace the group $SU(3)_{C} \times SU(2)_{L} \times U(1)_{Y}$ by a bigger one such as $SU(5)$ or $SO(10)$. One is then in $GUT$ (Grand Unified Theories). Such theories have their advantages and limitations. For example, the simplest $SU(5)$ theory has been shown to lead to a proton decay rate which is in contradiction with experiment . These theories introduce gauge bosons which are not present in the Standard Model. On the other side, these $GUT$ embed the possibility of explaining neutrino masses, especially using the See-saw Mechanism.\\

Another view is to keep the group $SU(3)_{C} \times SU(2)_{L} \times U(1)_{Y}$, the Lorentz group and introduce supersymmetry \cite{Wess:1992cp}\cite{West:1986wua}. The physical spectrum is thus doubled, with a superpartner field associated to each 
degree of freedom of the Standard Model.\\

Combining the two previous ideas, SUSY GUT can be constructed.\\

Other works have been devoted to the possibility of breaking Lorentz invariance. Some of their possible phenomenological implications have been analyzed.\\

Some approaches start with a modification of the usual commutation relations of quantum mechanics.\\

The work proposed here does not lie in the categories evoked above. The gauge group of the Standard Model as well as the Poincar\'e group are incorporated in the model. Our aim is to study some ideas coming from attemps to study quantum gravity \cite{Art04-01, Art04-02, Art04-03, Art04-04, Art04-05, Art04-06, Art04-07, Art04-08, Art04-09, Art04-10, Art04-01a, Art04-01b, Art04-01c, Art04-01d, Art04-01e, Art04-02a, Art04-02b, Art04-02c, Art04-02d, Art04-03a, Art04-03b}. Basically, if one has fields whose propagators are not the usual ones, that has influence on scattering cross sections,etc.\\

This paper is organized as follows. The first section is devoted to the "origins" of the fakeon hypothesis. It is based on attemps to study general relativity by introducing new terms to the Einstein-Hilbert action. Some of these extensions include terms which depend on derivatives of the Riemann tensor and thus lead to higher derivatives in the metric. \\

The second section deals with a simpler model relying on a real scalar field which displays the kind of behaviour needed by the fakeon hypothesis. Its Lagrangian involves high order derivatives of the field \cite{Art04-01, Art04-02, Art04-03, Art04-03c}. Its equations of motion are solved and show plane waves which, if interpreted naively in the usual sense, lead to a massless particle and a massive one. We verify that the equation of motion for the Green functions of this model behave in the desired way. We then make an heuristic consideration of a fakeon real scalar field which has a quartic self-interaction. Looking for a scattering involving two particles in the initial state and two in the final state, we see that the amplitude of transition at second order is finite, contrarily to the traditionnal, non fakeon treatment. Such a view has to be taken with great care. It basically relies on the replacement of the Feynman propagator of the usual theory by the one proposed by the fakeon hypothesis. A more detailed analysis is probably needed.\\

In the third section, we study the hypothesis that the Higgs field may be a fakeon, while all the other particles of the Standard Model are not. For the free fakeon Higgs doublet, we adapt the Standard Model Lagrangian to accomodate the gauge group of the electroweak theory. We then gauge this Lagrangian and find that it introduces new couplings between the physical Higgs and the electroweak gauge bosons.\\

The fourth section has a more formal and general accent. To approach the fakeon hypothesis, we used the simplest possible Lagrangian and it involves up to third order space-time derivatives of a scalar field. One can ask two simple questions. The first one is to know what happens if another particle of the Standard Model is also a fakeon. To study such a point of view, in which the fermions and or the gauge fields of the Standard Model could be fakeons, one has to consider Lagrangians which will bear ressemblances with what has been used here for the Higgs. We thus analyze formally a system in which the Lagrangian depends on a field and its derivatives 
up to the third order. We derive the equations of motion for such a system. We also derive its conserved quantities. We think this be useful for a deeper understanding of the hypothesis studied.\\

In the conclusions, we make a discussion concerning the interactions introduced by the fakeon hypothesis between the physical Higgs and the usual electroweak gauge bosons. And we give prospects
on how to test experimentally the nature of the Higgs boson, either a physical or a fake degree of freedom, through precision measurements at the next $e^{+}e^{-}$ collider envisaged to be in service right after the LHC era. \\

To make the reading of this article easier, we have put some technical computations in two appendices. The first one deals with the explicit calculation of the interaction terms introduced by the fakeon hypothesis. The second one is devoted to the derivation of the field equations and the conserved quantities (momenta, charges) for a general Lagrangian depending on up to third derivatives in the fundamental fields as given in section \ref{sec04}.\\

It should be clear that this work does not pretend to exhaust the approach of the fakeon hypothesis or of the virtual particle in Particle Physics as suggested for example in \cite{Art04-01, Art04-02, Art04-03}. The aim of this paper is quite modest. We do not study quantum gravity. We simply analyze how the fakeon hypothesis impacts a part of the electroweak theory, focusing on the Higgs field.

%%%%%%%%%%%%%%%%%%%%%%%%%%%%%%%%%%%%%%%%%%%%%%%%%%%%%%%%%%%%%%%%%%%%%%%%%%%%%%%%%%%%%%%%%%%%%%%%%%%%%%%%%%%%%%%%%%%%%%%%%%%%%
%%%%%%%%%%%%%%%%%%%%%%%%%%%%%%%%%%%%%%%%%%%%%%%%%%%%%%%%%%%%%%%%%%%%%%%%%%%%%%%%%%%%%%%%%%%%%%%%%%%%%%%%%%%%%%%%%%%%%%%%%%%%%

\section{Gravity and Higher-Derivatives of The Metric}
\label{sec02}

\ 

Theories with higher-derivatives in the metric field are present in some proposals for the extension of the Einstein-Hilbert Lagrangian describing gravity in a relativistic setting\cite{Art04-01}
\begin{eqnarray}
	\label{art4126}
	\begin{split}
		-2\kappa^2\mu^\epsilon\frac{\La_{GQ}}{\sqrt{-g}} & = 2\lambda_c M^2+\zeta R-\frac{\gamma}{M^2}R_{\mu\nu}R^{\mu\nu}+\frac{1}{2M^2}(\gamma-\eta)R^2\\
		&-\frac{1}{M^4}\left(D_\rho R_{\mu\nu}\right)\left(D^\rho R^{\mu\nu}\right)+\frac{1}{2M^4}(1-\xi)\left(D_\rho R\right)\left(D^\rho R\right)\\
		&\frac{1}{M^4}\left(\alpha_1R_{\mu\nu}R^{\mu\rho}R^\nu_\rho+\alpha_2RR_{\mu\nu}R^{\mu\nu}+\alpha_3R^3+\alpha_4RR_{\mu\nu\rho\sigma}R^{\mu\nu\rho\sigma}\right.\\
		&\qquad\qquad\left.+\alpha_5R_{\mu\nu\rho\sigma}R^{\mu\rho}R^{\nu\sigma}+\alpha_6R_{\mu\nu\rho\sigma}R^{\rho\sigma\alpha\beta}R^{\mu\nu}_{\alpha\beta}\right)
	\end{split}
\end{eqnarray}
where $\lambda_c, \zeta, \gamma, \eta, \xi, \alpha_i(1\leq i \leq 6)$ and $\kappa$ are constants.\\

The following expansion of the metric around the Minkowski one
\begin{equation}
	\label{art4127}
	g_{\mu\nu}=\eta_{\mu\nu}+2\kappa h_{\mu\nu}
\end{equation}
and the gauge-fixing function 
\begin{eqnarray}
	\label{art4128}
	\textbf{g}_\mu=\eta^{\nu\rho}\partial_\rho g_{\mu\nu}-\frac{1}{2}\eta^{\nu\rho}\partial_\mu g_{\rho\nu}.
\end{eqnarray}
can be used \cite{Art04-01}.\\
With all of the above, the gauge-fixed Lagrangian is:
\begin{eqnarray}
	\label{art4129}
	\begin{split}
		\La_{gf}&=\La_{GQ}+\frac{1}{4\kappa^2}\textbf{g}^\mu\left(\zeta-\gamma\frac{\Box}{M^2}+\frac{\Box^2}{M^4}\right)\textbf{g}_\mu\\ 
		&+\overline{C}^\mu\left(\zeta -\gamma\frac{\Box}{M^2}+\frac{\Box^2}{M^4}\right)\left[\Box C_\mu-(2\delta^\rho_\mu\eta^{\nu\sigma}\partial_\nu-\eta^{\rho\sigma}\partial_\mu)\Gamma^\alpha_{\rho\sigma}C_\alpha\right].	
	\end{split}
\end{eqnarray} 
The propagator has the following behavior
\begin{eqnarray}
	\label{art4130}
	i\tilde{G}_F(p)\sim\frac{1}{p^4} \ \  \text{for} \ p^2>>m^2.
\end{eqnarray}

%%%%%%%%%%%%%%%%%%%%%%%%%%%%%%%%%%%%%%%%%%%%%%%%%%%%%%%%%%%%%%%%%%%%%%%%%%%%%%%%%%%%%%%%%%%%%%%%%%%%%%%%%%%%%%%%%%%%%
%%%%%%%%%%%%%%%%%%%%%%%%%%%%%%%%%%%%%%%%%%%%%%%%%%%%%%%%%%%%%%%%%%%%%%%%%%%%%%%%%%%%%%%%%%%%%%%%%%%%%%%%%%%%%%%%%%%%%

\section{The Fakeon Hypothesis and Real Scalar Field}
\label{sec03}
\

The fakeon hypothesis relies on a modification of the propagators for fields needed to descibe fundamental particles in Physics. The simplest way to implement such an approach is to incorporate in the Lagrangian a supplementary term which contains a term with a third order derivative in the field. We took, from the literature, such a Lagrangian for a real scalar field.\\

With a real scalar field in hand, the simplest thing to do is to study a system in which such a field has quartic self- interactions and the easiest process is a scattering with two outgoing particles. Heuristically, one sees that the perturbative calculation of this scattering process does not lead, at the second order, to an ultraviolet divergence. This is in contrast with the usual treatment, which displays such a behaviour, taken care of by renormalization.\\

It turns out that for a non-interacting single scalar field $\phi$, one can write down a Lagrangian whose propagator displays interesting features for our purposes. It implies second derivatives in the field.\\
For the naive substitution : $\phi_1+\phi_2\longrightarrow \phi_3+\phi_4$, the second-order element $S_{fi}$ of the scattering matrix is non-divergent.\\

Consider the following Lagrangian
\begin{eqnarray}
	\label{art4131}
	\begin{split}
		\La&=\frac{1}{2}\left(\partial_\mu\varphi\right)\left(\zeta-\gamma\frac{\Box}{M^2}\right)\left(\partial^\mu\varphi\right)-\frac{\lambda}{4!}\varphi^4\\
		&=\frac{1}{2}\zeta \left(\partial_{\mu}\varphi\right)\left(\partial^{\mu}\varphi\right)-\frac{\gamma}{2 M^2}\left(\partial_{\mu}\varphi\right)\Box\left(\partial^{\mu}\varphi\right).
	\end{split}
\end{eqnarray}
The propagator takes the form
\begin{eqnarray}
	\label{art4132}
	\frac{i}{\zeta \left(p^2+i\epsilon\right)}-\frac{i\gamma\left(\zeta M^2+\gamma p^2\right)}{\zeta\left[\left(\zeta M^2+\gamma p^2\right)^2+\varepsilon^4\right]}
\end{eqnarray}
where $\epsilon$ et $\varepsilon$ are infinitesimal.\\
From the action
\begin{eqnarray*}
	\label{art4133}
	S=\int d^4x \ \La.
\end{eqnarray*}
one obtains
\begin{eqnarray}
	\begin{split}
	\label{art4134}
	\delta S &=\int d^4x\eta^{\mu\nu}\left\{ \zeta \partial_{\mu}\left(\partial_{\nu}\varphi \delta\varphi\right)-\frac{\gamma}{2M^2}\left[\partial_{\mu}\left(\delta\varphi\Box\partial_{\nu}\varphi\right)+\Box\partial_{\nu}\left(\partial_{\mu}\varphi\delta\varphi\right)\right]\right\}\\
	&+\int d^4x\eta^{\mu\nu}\left\{ -\zeta\partial_{\mu}\left(\partial_{\nu}\varphi\right)+\frac{\gamma}{2M^2}\left[\left(\partial_{\mu}\Box\partial_{\nu}\varphi\right)+\left(\Box\partial_{\nu}\partial_{\mu}\varphi\right)\right]\right\}\delta\varphi=0.
	\end{split}
\end{eqnarray} 
The equation of motion reads
\begin{equation}
	\label{art4135}
	\left(\zeta\Box +\frac{|\gamma|}{M^2}\Box^2\right)\varphi=0.
\end{equation}
Looking for plane wave solutions $\varphi= e^{ipx} $, equation \eqref{art4135} gives $\left(-\zeta p^2+\frac{|\gamma|}{M^2}p^4\right)+0.$ This implies 
\begin{itemize}
	\item either 
	\begin{equation}
		\label{art4136}
		p^2=0\Longrightarrow p^2_0=\vec{p}^{2} \Longrightarrow m^2=0,
	\end{equation}  
	\item or 
	\begin{equation}
		\label{art4137}
		\frac{|\gamma|}{M^2}p^2=\zeta
	\end{equation}
\end{itemize}

\begin{align}
	\left(\Box_x+m^2\right)G_{Fs}(x-y) &= i\delta^4(x-y) \ \text{et} \label{art493a}\\
	\left(\zeta\Box_x +\frac{|\gamma|}{M^2}\Box^2_x\right)G_{Ff}(x-y) &= i\delta^4(x-y).\label{art493b}
\end{align}

\begin{align}
iG_{Fs}(x-y) &= \int \frac{d^4k}{(2\pi)^4}\frac{1}{k^2-m^2+i\epsilon}e^{-ik(x-y)} \ et \label{art494a}\\
iG_{Ff}(x-y) &= \int \frac{d^4k}{(2\pi)^4}\frac{M^2}{k^2\left(\zeta M^2  -|\gamma|k^2\right)+\varepsilon^4}e^{-ik(x-y)}.\label{art494b}
\end{align}
with $\epsilon$ and $\varepsilon$ infinitesimal widths.

\begin{equation}
	\label{art495}
	p_1+p_2 \longrightarrow p_3+p_4
\end{equation}

\begin{align}
\ket{i} &=a^\dagger_{\vec{p}_1}a^\dagger_{\vec{p}_2}\ket{0}\qquad\mbox{et}\qquad \label{art496a} \\ 
\ket{f} &=a^\dagger_{\vec{p}_3}a^\dagger_{\vec{p}_4}\ket{0}. \label{art496b}
\end{align}

\begin{enumerate}
	\item {\bf Order 0}\\
	This term does not describe an interaction.
	
	\item {\bf First Order}\\
	\begin{equation}
		\label{art497}
		S_{fi}^1 = -i\lambda(2\pi)^4\delta^4\left(p_4+p_3-p_1-p_2\right).
	\end{equation}
	
	\item {\bf Second Order}\\
	\begin{eqnarray}
		\label{art498}
		\begin{split}
		S_{fi}^2&=\frac{-\lambda^2}{2}\delta^4(p_3+p_4-p_1-p_2)\int d^4q \left[\frac{1}{q^2-m^2+i\epsilon}\frac{1}{\left(p_1+p_2-q\right)^2-m^2+i\epsilon}\right.\\
		&\left.+\frac{1}{q^2-m^2+i\epsilon}\frac{1}{\left(p_2-p_4-q\right)^2-m^2+i\epsilon}+\frac{1}{q^2-m^2+i\epsilon}\frac{1}{\left(p_2-p_3-q\right)^2-m^2+i\epsilon}\right].
		\end{split}	
	\end{eqnarray}

	\begin{equation}
		\label{art499}
		S_{fi}^2\propto\int_{|q|\rightarrow+\infty} \frac{d^4q}{q^4}\propto\int_{|q|\rightarrow+\infty}  \frac{q^3}{q^4}dq=\int_{|q|\rightarrow+\infty}  \frac{dq}{q}=[lnq]_{|q|\rightarrow+\infty}\longrightarrow\infty.
	\end{equation}

	\item {\bf Amplitude $S_{fi}^2$ for fakeons}\\
	\begin{eqnarray}
		\label{art4100}
		\begin{split}
		S_{fi}^2 &= \frac{-\lambda^2}{2}\delta^4(p_3+p_4-p_1-p_2)\\
				& \times \int d^4q \left[\frac{M^2}{q^2\left(\zeta M^2  -|\gamma|q^2\right)+\varepsilon^4}\right. \frac{M^2}{ \left(p_1+p_2-q\right)^2\left[ \zeta M^2-|\gamma|\left(p_1+p_2-q\right)^2\right]+\varepsilon^4}\\
				&+\frac{M^2}{q^2\left(\zeta M^2  -|\gamma|q^2\right)+\varepsilon^4}\frac{M^2}{ \left(p_2-p_4-q\right)^2\left[\zeta M^2 -|\gamma|\left(p_2-p_4-q\right)^2\right]+\varepsilon^4}\\
				&\left.+\frac{M^2}{q^2\left(\zeta M^2  -|\gamma|q^2\right)+\varepsilon^4}\frac{1}{ \left(p_2-p_3-q\right)^2\left[\zeta M^2 -|\gamma|\left(p_2-p_3-q\right)^2\right]+\varepsilon^4}\right].\label{eq88}
		\end{split}
	\end{eqnarray}

	\begin{equation}
		\label{art4138}
		S_{fi}^2\propto\int_{|q|\rightarrow+\infty} \frac{d^4q}{q^8}\propto\int_{|q|\rightarrow+\infty}  \frac{q^3}{q^8}dq=\int_{|q|\rightarrow+\infty} \frac{dq}{q^5}=\left[\frac{-1}{4q^4}\right]_{|q|\rightarrow+\infty}\longrightarrow 0.
	\end{equation}
\end{enumerate}

Let us now look at the simplest process in this model i.e. a scattering one.

\begin{equation}
\label{art4155}
\phi\left(\overrightarrow{p}_1\right) + \phi\left(\overrightarrow{p}_2\right) \longrightarrow \phi\left(\overrightarrow{p}_3\right) \phi\left(\overrightarrow{p}_4\right).
\end{equation}
Knowing the initial and final states, quantum field theory tell us that the amplitude of the process is given by $\braket{f \vert S \vert i}$ where $S=T \exp{\Big[ \dfrac{-i}{\hbar} \displaystyle\int d^4 x \ \mathcal{L}_{\text{int}} (x)  \Big]}$. Using the decomposition of the free field
\begin{equation}
\label{art4165}
\phi(x)=\int d^3 k \ \Big[ a(\overrightarrow{k}) e^{-i k.x} + a^\dagger(\overrightarrow{k}) e^{i k.x} \Big],
\end{equation}
the commutation relations are $\Big[ a(\overrightarrow{k}) , a^\dagger(\overrightarrow{k}') \Big]=\delta^3 \left( \overrightarrow{k} - \overrightarrow{k}'\right); \ \Big[ a(\overrightarrow{k}) , a(\overrightarrow{k}') \Big]=\Big[ a^\dagger(\overrightarrow{k}) , a^\dagger(\overrightarrow{k}') \Big]=0$.\\

The non-fakeon  treatment leads to Eq.\eqref{art497}, Eq.\eqref{art498} where the integration is written in terms of the Feynman propagators for the scalar field i.e. $\dfrac{1}{p^2-m^2+i\varepsilon}$. This integral is divergent and can be taken of by renormalization, in the ultraviolet.\\

If one takes the option of simply replacing the usual propagator $\dfrac{1}{p^2-m^2+i\varepsilon}$ by another one $\dfrac{p^2-m^2}{\left(p^2-m^2+i\varepsilon\right)^2+\epsilon}$\cite{Art04-01, Art04-02, Art04-03}, one find that amplitude of the scattering process is finite at the second order contrary to what happens usually.\\

The important point is that the Feynman propagator appearing here is given as $G_F(x-y)=\braket{0\vert T \phi(x) \phi(y) \vert0}$ where the field $\phi$ has been quantized, according to the rules invoked above.\\

We have found the classical solutions of the field equation (See Eq.\eqref{art4165}) but have not provided the usual quantization. This means that claim that the $S_{\text{fi}}^{(2)}$ not being divergent has to be examined more closely. Section \ref{sec04} gives a first step for the solution of this problem.

%%%%%%%%%%%%%%%%%%%%%%%%%%%%%%%%%%%%%%%%%%%%%%%%%%%%%%%%%%%%%%%%%%%%%%%%%%%%%%%%%%%%%%%%%%%%%%%%%%%%%%%%%%%%%%%%%%%%%
%%%%%%%%%%%%%%%%%%%%%%%%%%%%%%%%%%%%%%%%%%%%%%%%%%%%%%%%%%%%%%%%%%%%%%%%%%%%%%%%%%%%%%%%%%%%%%%%%%%%%%%%%%%%%%%%%%%%%

\section{The Higgs Boson As a Fakeon}
\label{sec04}
\

In this paper, we begun with a Lagrangian describing a real scalar field, with the necessary ingredients to accomodate the fakeon hypothesis. We then promoted that single real scalar field to a doublet of complex fields needed for the Higgs sector in the Standard Model. The Lagrangian obtained in that fashion is invariant under global transformations of the Standard Model. We then went on to make that symmetry local, using the usual recipe of replacing the partial space-time derivative by the covariant derivative.\\

The presence of the D'Alembertian operator in the Lagrangian leads to some lengthy expressions for the interaction between the Higgs boson and the electroweak gauge bosons.
\begin{equation}
	\label{art4101}
	\mathcal{L}(\Phi) = \xi \left(D_\mu \Phi\right)^\dagger \ D^\mu \left(D_\nu D^\nu \Phi\right) + \left(D_\mu \Phi\right)^\dagger \left(D^\mu \Phi\right).
\end{equation}
Since the covariant derivative commutes with the gauge transformation i.e.
\begin{equation}
	\label{art4102}
	\Phi = U \ \Phi'; \ D_\mu \Phi= U \ D_\mu \Phi'
\end{equation}
for the group $SU(2)_{L}$, one has
\begin{equation}
	\label{art4103}
	\mathcal{L}(\Phi') = \xi \left(D_\mu \Phi'\right)^\dagger \ U^\dagger U \ D^\mu \left(D_\nu D^\nu \Phi'\right) + \left(D_\mu \Phi'\right)^\dagger \ U^\dagger U \ \left(D^\mu \Phi'\right)
\end{equation}
so that the Lagrangian is invariant when the matrix $U$ belongs to $SU(2)_{L}$ i.e. $\mathcal{L}(\Phi)=\mathcal{L}(\Phi')$. The same can be said of the $U(1)_{Y}$ part of the electroweak theory.\\

Gauging the theory gives extra couplings of the electroweak bosons $\left(W^+_\mu, W_\mu^-, Z_\mu, A_\mu\right)$ to the physical spin zero Higgs field $h$ after spontaneous symmetry breaking.\\
\begin{align}
	W^\pm_\mu &= \frac{1}{\sqrt{2}}\left(A^1_\mu\mp iA^2_\mu\right);\label{art479a}\\
	Z_\mu &= \frac{1}{\sqrt{g^2+g'^2}}\left(gA^3_\mu-g'B_\mu\right);\label{art479b}\\
	A_\mu &= \frac{1}{\sqrt{g^2+g'^2}}\left(g'A^3_\mu+gB_\mu\right).\label{art479c}
\end{align}

\begin{equation}
	\label{art484}
	A^1_\mu = \frac{1}{\sqrt{2}}\left(W^+_\mu+W^-_\mu\right);
\end{equation}

\begin{equation}
	\label{art486}
	A^2_\mu = \frac{i}{\sqrt{2}}\left(W^+_\mu-W^-_\mu\right);
\end{equation}

\begin{equation}
	\label{art488}
	A^3_\mu = \frac{1}{\sqrt{g^2+g'^2}}\left(gZ_\mu+g'A_\mu\right);
\end{equation}

\begin{equation}
	\label{art490}
	B_\mu = \frac{-1}{\sqrt{g^2+g'^2}}\left(g'Z_\mu-gA_\mu\right).
\end{equation}

\begin{eqnarray}
	\label{art491}
	\begin{split}
	(D_\mu \Phi)^\dagger(D^\mu \Phi)&=&=\frac{1}{2}\partial_\mu h\partial^\mu h +\left(v+h\right)^2\frac{g^2}{4}W^-_\mu W^{+\mu}+\left(v+h\right)^2\frac{\left(g^2+g'^2\right)}{8}Z_\mu Z^\mu.
	\end{split}
\end{eqnarray}

\begin{eqnarray}
	\label{art492}
	\begin{split}
	\left[\Phi^\dagger\Phi-\frac{v^2}{2}\right]^2=&\frac{1}{4}\left(h^2+2vh\right)^2.
	\end{split}
\end{eqnarray}
The covariant derivative of the Higgs doublet can be written in the following form which simplifies computations:
\begin{equation}
	\label{art4139}
	D_\mu \Phi= \partial_\mu\Phi+\sum_{a=1}^{4} K_{a,\mu}\tilde{T}_a\Phi
\end{equation}
where one takes 
\begin{equation}
K_{1,\mu}= W^{+}_{\mu}, \ K_{2,\mu}= W^{-}_{\mu}, \ K_{3,\mu}= Z_{\mu}, \ K_{4,\mu}= A_{\mu}
\end{equation}
so that
\begin{equation}
\tilde{T}_1 = \left(
\begin{array}{cc}
 0 & -\dfrac{i g}{\sqrt{2}} \\
 0 & 0
\end{array}
\right), \
\tilde{T}_2 = \left(
\begin{array}{cc}
 0 & 0 \\
 -\dfrac{i g}{\sqrt{2}} & 0
\end{array}
\right)
\end{equation}

\begin{equation}
\tilde{T}_3 =\left(
\begin{array}{cc}
 -i g \left(\dfrac{1}{2}-\sin ^2(\theta )\right) \sec (\theta ) & 0 \\
 0 & \dfrac{1}{2} i g \sec (\theta )
\end{array}
\right), \
\tilde{T}_4 =\left(
\begin{array}{cc}
 -i e & 0 \\
 0 & 0
\end{array}
\right).
\end{equation}
The part of the Lagrangian containig the D'Alembertian can be expanded directly in term of the physical electroweak gauge bosons. An example of a new interaction is given by  
\begin{eqnarray}
	\label{art4140}
	\begin{split}
		\La_2 &= \sum_{a=1}^{4} \left( \partial^{\mu} \partial_{\nu} K_a, ^\nu \right) \ \dfrac{1}{2} \partial_{\mu}\Phi^\dagger \left(v+h\right) \ \dfrac{ig}{2} \sec\theta_W \ \delta_{a, 3}\\
		&= \dfrac{i}{4} g \sec\theta_W \left( \partial^{\mu} \partial_{\nu} K_3, ^\nu \right) \ \left(\partial_{\mu} h\right) \left(v+h\right)\\
		\La_2 &=\dfrac{i}{4} g \sec\theta_W \left( \partial^{\mu} \partial_{\nu} Z^\nu \right) \ \left(\partial_{\mu} h\right) \left(v+h\right).
	\end{split}
\end{eqnarray}
In the same way, we calculate the other terms, we have then :
\begin{eqnarray}
	\label{art4141}
	\begin{split}
	\La =&\quad(\partial_\mu h)(\partial^\mu \Box h)\\
	&+\frac{1}{4}ig\sec\theta_W\partial_\mu h(v+h)\left(\partial^\mu\partial_\nu Z^\nu\right)\\
	&+\frac{1}{4}ig\sec\theta_W(\partial_\mu h)(\partial^\mu h)\partial_\nu Z^\nu\\
	&+ig\sec\theta_W(\partial_\mu h) (\partial_\nu h)\left(\partial^\mu Z^\nu\right)\\
	&+ig \sec\theta_W Z^\nu(\partial_\mu h )(\partial^\mu \partial_\nu h)\\
	&-\frac{1}{4}ig \sec\theta_W (\partial^\mu\Box h)(v+h)Z_\mu \\
	&-\frac{1}{4}g^2(\partial_\mu h) (v+h) \left[\left(\partial^\mu W^-_\nu\right)W^{+\nu}+\frac{1}{2} \sec^2\theta_W\left(\partial^\mu Z_\nu\right)Z^\nu\right]\\
	&-\frac{1}{2}g^2(\partial_\mu h) (v+h)\left[\left(\partial^\mu W^{+\nu} \right)W^-_\nu+\frac{1}{2} \sec^2\theta_W\left(\partial^\mu Z^\nu\right)Z_\nu\right]\\
	&-\frac{1}{4}g^2(\partial_\mu h)(\partial^\mu h) \left[ W^{+\nu} W^-_\nu+\frac{1}{2} \sec^2\theta_W Z^\nu Z_\nu\right]\\
	&-\frac{1}{4}g^2 (v+h)^2\left[\left(\partial^\mu\partial_\nu W^{+\nu} \right)W^-_\mu+\frac{1}{2} \sec^2\theta_W\left(\partial^\mu \partial_\nu Z^\nu\right)Z_\mu\right]\\
	&+ ...
	\end{split}
\end{eqnarray}
The details of the computations are given in Appendix \ref{App01}.\\

%Before the discovery of the Higgs boson at the LHC, the mass of the boson $m_h$ was a free parameter. The other parameters of the Standard Model are $m_b=5GeV, m_t=175 GeV, m_W=80 GeV, m_Z=91 GeV, \sin^2\theta_W=0.23, \alpha_S(m_Z)=0.12$.\\

The Standard Model displays explicitly the couplings of the Higgs field with the fermions and gauge bosons involved. Although the mass of the Higgs was unknow til $2012$, the picture was already clear and constrained by the fact that the interaction of the Higgs boson with a fermion is linked to the mass of spin half particle via the Yukawa terms in the Lagrangian. On the other side, the masses of the electroweak gauge bosons can be related to the coupling constants of the model $g, g'$, the hypercharge $Y$ of the Higgs doublet and its vacuum expectation value $v$.\\

Basically, the search of the Higgs boson was based on its decays rates and its production rate. The part of the Lagrangian implying the Higgs field, after spontaneous symmetry breaking reads
\begin{equation}
	\label{art4193}
	\mathcal{L}_{\text{Higgs}}=\dfrac{1}{2} \partial_\mu h \ \partial^\mu h - \dfrac{\lambda}{4} \left(h^4 + 4 v^2 h^2 + 4 v h^3\right) + \dfrac{1}{8} \left(v+h\right)^2 \Big[ 2 g^2 W^+_\mu W^{-\mu} + \left(g^2+g'^2\right) Z_\mu Z^\mu\Big]
\end{equation}
with $m_h=\sqrt{2\lambda} v$.\\

The main decays rates are given by\cite{Art04-15}:
\begin{enumerate}[label=\alph*]
	\item \textbf{The decay mode into two fermions}\\
	
	The rate is found to be
	\begin{equation}
		\label{art4194}
		\Gamma(h\longrightarrow f\overline{f}) = \left(\dfrac{\alpha m_h}{8\sin^2\theta_W}\right)\dfrac{m_f^2}{m^2_W}\left(1-\dfrac{4 m_f^2}{m^2_h}\right)^{3/2} N_c(f)
	\end{equation}
	with $\alpha=\dfrac{1}{137}, N_c(f)=1$ for leptons, $N_c(f)=3$, for quarks and $2 m_f \leq m_h$ (this excludes the top quark).
	
	\item \textbf{The decay mode into two gluons}\\
	
	\begin{equation}
		\label{art4195}
		\Gamma(h\longrightarrow 2g) = \left(\dfrac{\alpha m_h}{8\sin^2\theta_W}\right)\dfrac{m_f^2}{m^2_W}. \dfrac{\alpha_S}{9 \pi^2} \Big\vert \sum_q I\left(\dfrac{m_h^2}{m_q^2}\right)\Big\vert^2
	\end{equation}
	with $I(x)$ a form factor. Among its properties, one has $\lim\limits_{x\longrightarrow0} I(x)=1$ and $\lim\limits_{x\longrightarrow\infty} I(x)=0$. As a consequence, the heavy quarks contribute more than the light ones to this process. The inverse of this process can be seen as a precessus creating Higgs from two gluons.
	
	\item \textbf{The decay mode into two photons}\\
	
	\begin{equation}
	\label{art4196}
	\Gamma(h\longrightarrow 2\gamma) = \left(\dfrac{\alpha m_h}{8\sin^2\theta_W}\right)\dfrac{m_f^2}{m^2_W}. \dfrac{\alpha_S}{18 \pi^2} \Big\vert \sum_F Q^2_f N_c(f)-\dfrac{21}{4}\Big\vert^2.
	\end{equation}
	We didn't include the point of view of \cite{Art04-15} $m_h>2 m_W$, since it is excluded by LHC.
\end{enumerate}
\
Going beyond the Standard Model does not mean one is free to do whatever one wants. The discovery of the Higgs boson poses stricts constraints in this matter.\\

That fakeon hypothesis, in the minimal setting studied here, implies a modification of the Higgs propagator and new interactions with a common factor $\xi$:
\begin{equation}
	\label{art4197}
	\mathcal{L}_{\text{int}}=\xi \left(\mathcal{L}_1 + \mathcal{L}_2 + ... + \mathcal{L}_{24} \right).
\end{equation}
The fakeon hypothesis led us to introduce the term $\left(D_\mu \Phi\right)^\dagger \ D^\mu \left(D_\nu D^\nu \Phi \right)$. It should be noted that this quantity is not always a real quantity. That led us to introduce the quantity
\begin{equation}
	\label{art4198}
	\mathcal{L}_\star=\left(D_\mu \Phi\right)^\dagger \ D^\mu \left(D_\nu D^\nu \Phi \right) + h.c.
\end{equation}
which has the proper behaviour.\\

The equations \eqref{art4194}, \eqref{art4195} and \eqref{art4196}, tested experimentally, are in accord with the theoretical body of the Standard Model.\\

The fakeon hypothesis, as treated here, introduces a new constant and some non trivial interactions.\\

It is to be seen how that hypothesis changes the predictions of the rates given in \eqref{art4194}, \eqref{art4195}, \eqref{art4196}. In principle, those rates will depend on the new parameter $\gamma'$, with $\gamma'=0$ giving simply the Standard Model. Such that the quantity $\gamma'$ has to be very small $\vert \gamma' \vert <<< 1$.

%%%%%%%%%%%%%%%%%%%%%%%%%%%%%%%%%%%%%%%%%%%%%%%%%%%%%%%%%%%%%%%%%%%%%%%%%%%%%%%%%%%%%%%%%%%%%%%%%%%%%%%%%%%%%%%%%
%%%%%%%%%%%%%%%%%%%%%%%%%%%%%%%%%%%%%%%%%%%%%%%%%%%%%%%%%%%%%%%%%%%%%%%%%%%%%%%%%%%%%%%%%%%%%%%%%%%%%%%%%%%%%%%%%

\section{Equations of Motion and Conserved Quantities}
\label{sec05}

\begin{equation}
	\label{art4142}
	\varPi=\frac{\partial\La}{\partial(\partial_0\varphi)}.
\end{equation}

\begin{equation}
	\label{art412}
	\left[\varphi \left(\vec{x},t\right), \varPi\left(\vec{y},t\right)\right]=\delta^3\left(\vec{x}-\vec{y}\right).
\end{equation}

\begin{equation}
	\label{art409}
	\La=\textbf{A}\left(\partial_{\mu}\varphi\right)\left(\partial^{\mu}\varphi\right)+\textbf{B}\varphi^2+\textbf{C}\left(\partial_{\mu}\varphi\right)\Box\left(\partial^{\mu}\varphi\right).
\end{equation}

To understand the complete physical meaning of the Lagrangian given by \eqref{art409} (with C=0), it is necessary to associate to its conserved quantities as indicated by the Noether theorem. The invariance of the Lagrangian under the Poincar\'e group leads to the following conserved quantities: 
\begin{enumerate}
	\item The moments $P_\mu$, related to the invariance under translations;
	\item The "generalized" angular momentum $M_{\mu\nu}$, related to the Lorentz transformations. When $\mu$ and $\nu$ are spatial indices, it is simply the angular momentum.
\end{enumerate}
These conserved quantities, at the classical level, obey the following relations, which give the representation of the Poincar\'e group.
\begin{align}
\Big[ P_\mu , P_\nu \Big] &=0; \label{art413a}\\
\Big[ M_{\mu\nu} , P_\tau \Big] &=\eta_{\mu\tau}P_\nu-\eta_{\nu\tau} P_\mu;\label{art413b}\\
\Big[ M_{\mu\nu} , M_{\sigma\rho} \Big] &=\eta_{\nu\sigma}M_{\mu\rho}+\eta_{\mu\rho}M_{\nu\sigma}-\eta_{\nu\rho}M_{\mu\sigma}-\eta_{\mu\sigma}M_{\nu\rho}.\label{art413c}
\end{align}
At the quantum level, when the $\varphi$ field and its conjugate $\varPi$ have been promoted to the rank of quantum operators. These obey the same relation as the one given by the expression \eqref{art412} to the nearest $i\hbar$ (in the natural unit system: $\hbar=c=1$).
The dilemma we are facing is the following: the simplest Lagrangian involving the \textit{fakeons} is based on a derivative of the third order field. The equations of motion are of order 4. The Lagrangian is of the form (when $C\neq 0$):
\[\La\left(\varphi, \partial_\mu\varphi, \partial_\mu\partial_\mu\varphi, \partial_\mu\partial_\mu\varphi, \partial_\mu\partial_\mu\varphi\right).\]
This is different from the usual framework where only the first derivatives of the field appear. When $C\not=0$, the $\varphi$ field has more degrees of freedom and therefore the commutation relation given in \eqref{art412} does not "settle" everything.
The procedure we choose is to solve the equation of motion for $C\not=0$. The general solution will depend on some arbitrary constants. The conserved quantities coming from the Poincar\'e symmetry are also derived.

\begin{eqnarray}
	\label{art4143}
	\left\lbrace\begin{array}{ll}
	x^{\mu}\longrightarrow x'^{\mu}=x^\mu+\delta x^\mu\\
	\varphi_r(x)\longrightarrow\varphi'_r(x')=\varphi_r(x)+\delta\varphi_r(x)
	\end{array}\right. .
\end{eqnarray}
Let
\begin{equation}
	\label{art4144}
	\delta\varphi_r=\varphi'_r(x')-\varphi_r(x).
\end{equation}

\begin{equation}
	\label{art4145}
	\delta\varphi_r(x)=\tilde{\delta}\varphi_r(x)+\frac{\partial\varphi_r(x)}{\partial x^\mu}\delta x^\mu.
\end{equation}

\begin{equation}
	\label{art4146}
	\delta\varphi_r=\
	\partial_\mu\delta\varphi_r(x)=\frac{\partial\varphi_r(x)}{\partial x^\nu}\frac{\partial\delta x^\nu}{\partial x^\mu}+\delta\left(\partial_\mu\varphi_r(x)\right).
\end{equation}

\begin{eqnarray}
	\label{art4147}
	\begin{split}
	\delta S=0=&\int d^4x\left[\frac{\partial\La}{\partial\varphi_r(x)}\tilde{\delta}\varphi_r(x)+\frac{\partial\La}{\partial\partial_\mu\varphi_r(x)}\tilde{\delta}\partial_\mu\varphi_r(x)+\frac{\partial\La}{\partial\left(\partial_\mu\partial_\nu\varphi_r(x)\right)}\tilde{\delta}\left(\partial_\mu\partial_\nu\varphi_r(x)\right)\right.\\
	&\left.+\frac{\partial\La}{\partial\left(\partial_\mu\partial_\nu\partial_\sigma\varphi_r(x)\right)}\tilde{\delta}\left(\partial_\mu\partial_\nu\partial_\sigma\varphi_r(x)\right)+\partial_\mu\La\delta x^\mu +\La\partial_\mu\delta x^\mu\right].
	\end{split}
\end{eqnarray}

\begin{equation}
	\label{art4148}
	\delta S=\int d^4x\left[\Omega^r\tilde{\delta}\varphi_r +\partial_\mu f^\mu\right].
\end{equation}

\begin{equation}
	\label{art4149}
	\frac{\partial\La}{\partial\varphi_r}- \partial_\mu\left(\frac{\partial\La}{\partial\left(\partial_\mu\varphi_r\right)}\right)+\partial_\mu\partial_\nu\left(\frac{\partial\La}{\partial\left(\partial_\mu\partial_\nu\varphi_r(x)\right)}\right)-\partial_\mu\partial_\nu\partial_\tau\left(\frac{\partial\La}{\partial\left(\partial_\mu\partial_\nu\partial_\tau\varphi_r(x)\right)}\right)=0.
\end{equation}

\begin{eqnarray}
	\label{art4150}
	\Theta^\mu_\alpha =& P^{r,\mu}\partial_\alpha\varphi_r + R^{r,\mu\nu}    \partial_\nu\partial_\alpha\varphi_r+C^{r,\mu\nu\tau}\partial_\nu\partial_\tau\partial_\alpha\varphi_r-\eta^\mu_\alpha\La.
\end{eqnarray}

\begin{eqnarray}
	\label{art4151}
	P_\alpha=\int_V d^3x \ \Theta^0_\alpha.
\end{eqnarray}

\begin{equation}
	\label{art4152}
	\Lambda^\mu_\nu=\delta^\mu_\nu +\eta^{\mu\rho}\delta\omega_{\rho\nu}.
\end{equation}

%%%%%%%%%%%%%%%%%%%%%%%%%%%%%%%%%%%%%%%%%%%%%%%%%%%%%%%%%%%%%%%%%%%%%%%%%%%%%%%%%%%%%%%%%%%%%%%%%%%%%%%%%
%%%%%%%%%%%%%%%%%%%%%%%%%%%%%%%%%%%%%%%%%%%%%%%%%%%%%%%%%%%%%%%%%%%%%%%%%%%%%%%%%%%%%%%%%%%%%%%%%%%%%%%%%

\section{Conclusion}
\label{sec06}
\
The question of how the fakeon hypothesis can be implemented in Particle Physics has already led to many works \cite{Art04-01, Art04-02, Art04-03, Art04-03c}. Some aspects have been considered in relation to renormalizability and quantum gravity.\\

In this work, we analysed the hypothesis of the Higgs field of the Standard Model being a fakeon. For this, we relied on the simplest lagrangian which produces a propagator with the desired behaviour for a real scalar field. This Lagrangian involves up to third derivatives in the field.\\

After that, we proceeded to consider a doublet of complex scalar fields with a similar Lagrangian. We then went on to make that symmetry local, using the standard replacement of the partial derivative $\partial_\mu$ by a covariant derivative $D_\mu$.\\

These new interactions have of course to be added to those already present in the standard electroweak theory. Each of them can be written as a product of two factors. In the products, he first factors are polynomial in the gauge bosons fields and its derivatives, and the second ones are polynomial in the expression of these interactions.\\

The second part of our work dealt with the hypothesis that of all the particles of the Standard Model, only the Higgs was a fakeon. The Higgs boson is the last particle to have been detected at the CERN LHC experiments, ATLAS and CMS, in accordance with the Standard Model. Many extensions of the Standard Model exist. The fakeon hypothesis can be considered as one of them. Future experiments (FCC-ee or ILC or CLIC) will tell if such an approach is relevant \cite{Art04-11, Art04-12, Art04-13}.\\

The initial Lagrangian contains third order derivatives of the scalar field. This leads to an interaction Lagrangian containing many components. However, each of them can be written as a product of two factors. The first such factors depend solely on the electroweak bosons $\left(W^+, W^-, Z, A\right)$. The second ones only on the Higgs field and its derivatives.\\

The first terms evoked above are polynomials in the electroweak gauge bosons and their derivatives. The degree of these polynomials goes from one to four while the derivative order goes from zero to two. The same happens for the Higgs field. The corresponding polynomials are all quadratic in terms of degree. The derivative order goes from zero to three.\\
These two statements can be considered as the main result of this work.\\

Of course, this work is only a first step. One needs to study in detail how the strength of the interactions deduced from the fakeon hypothesis can be restricted by the available experimental data. Another, natural step should be to implement the fakeon hypothesis for the other particles of the Standard Model. There is a priori no fundamental reason for the Higgs boson to behave in such a different manner compared to the other particles. The last point is more formal. It is related to the general Lagrangian needed for the fakeon hypothesis and the associated conserved quantities. And not only for the scalar field. We are planing to adress these questions in a near future.\\

%%%%%%%%%%%%%%%%%%%%%%%%%%%%%%%%%%%%%%%%%%%%%%%%%%%%%%%%%%%%%%%%%%%%%%%%%%%%%%%%%%%%%%%%%%%%%%%%%%%%%%
%%%%%%%%%%%%%%%%%%%%%%%%%%%%%%%%%%%%%%%%%%%%%%%%%%%%%%%%%%%%%%%%%%%%%%%%%%%%%%%%%%%%%%%%%%%%%%%%%%%%%%

\underline{Acknowledgement}\\

The authors of this paper are greatful for the financial support provided by the CNRS/DERCI (France). Our project was supported under their IEA (International Emergency Action) and their DSCA (Dispositif de Soutien aux Collaborations avec l'Afrique sub-saharienne) programs.
 
%%%%%%%%%%%%%%%%%%%%%%%%%%%%%%%%%%%%%%%%%%%%%%%%%%%%%%%%%%%%%%%%%%%%%%%%%%%%%%%%%%%%%%%%%%%%%%%%%%%%%%
%%%%%%%%%%%%%%%%%%%%%%%%%%%%%%%%%%%%%%%%%%%%%%%%%%%%%%%%%%%%%%%%%%%%%%%%%%%%%%%%%%%%%%%%%%%%%%%%%%%%%%

\pagebreak

%%%%%%%%%%%%%%%%%%%%%%%%%%%%%%%%%%%%%%%%%%%%%%%%%%%%%%%%%%%%%%%%%%%%%%%%%%%%%%%%%%%%%%%%%%%%%%%%%%%%%%%
%%%%%%%%%%%%%%%%%%%%%%%%%%%%%%%%%%%%%%%%%%%%%%%%%%%%%%%%%%%%%%%%%%%%%%%%%%%%%%%%%%%%%%%%%%%%%%%%%%%%%%%
%%%%%%%%%%%%%%%%%%%%%%%%%%%%%%%%%%%%%%%%%%%%%%%%%%%%%%%%%%%%%%%%%%%%%%%%%%%%%%%%%%%%%%%%%%%%%%%%%%%%%%%
	
\appendix

\pagebreak

\section{New Interactions Between The Physical Higgs and The Electroweak Gauge Bosons}
\label{App01}

\

In this part of the paper, we expand the Lagrangian needed for the fakeon Higgs, dropping the common constant $\gamma'$ which multiplies it:
\begin{equation}
	\label{art4192}
	\mathcal{L}=\left(D_\mu \Phi\right)^\dagger \ D^\mu \left(D_\nu D^\nu \Phi \right)
\end{equation}
with the covariant derivative $D_\mu$ given by Eq.\eqref{art4139}. The Lagrangian \eqref{art4192} has four derivatives of the Higgs field. Using the covariant derivative \eqref{art4139}, the Lagrangian terms of \eqref{art4192} are given below. Some these terms purely are imaginary. In the final development of the Lagrangian \eqref{art4192}, these imaginary terms fall off, because we add the conjugate complex of the terms. Here are the different terms of the Lagrangian above:
\begin{align}
\La_1 &= (\partial_\mu \Phi^\dagger)(\partial^\mu \Box\Phi); \label{art4102a} \\
\La_2 &= \sum_{a=1}^{4} \left(\partial_\mu \Phi\right)^\dagger \ \left(\partial^\mu\partial_\nu K_{a,}^\nu\right)\tilde{T}_a \ \Phi; \label{art4102b} \\
\La_3 &= \sum_{a=1}^{4} \left(\partial_\mu \Phi\right)^\dagger \ \left(\partial_\nu K_{a,}^\nu\right)\tilde{T}_a \ \left(\partial^\mu\Phi\right); \label{art4102c} \\
\La_4 &= 2\sum_{a=1}^{4} \left(\partial_\mu \Phi\right)^\dagger \ \left(\partial^\mu K_{a,}^\nu\right)\tilde{T}_a \ \left(\partial_\nu\Phi\right); \label{art4102d} \\
\La_5 &= 2\sum_{a=1}^{4} \left(\partial_\mu \Phi\right)^\dagger \ K_{a,}^\nu \tilde{T}_a \ \left(\partial^\mu\partial_\nu\Phi\right); \label{art4102e} \\
\La_6 &= \sum_{a,b=1}^{4} \left(\partial_\mu \Phi\right)^\dagger \ \left(\partial^\mu K_{a,\nu}\right) K_{b,}^\nu \tilde{T}_a\tilde{T}_b \ \Phi; \label{art4102f} \\
\La_7 &= \sum_{a,b=1}^{4} \left(\partial_\mu \Phi\right)^\dagger \ K_{a,\nu} \left(\partial^\mu K_{b,}^\nu\right) \tilde{T}_a\tilde{T}_b \ \Phi; \label{art4102g} \\
\La_8 &= \sum_{a,b=1}^{4} \left(\partial_\mu \Phi\right)^\dagger \ K_{a,\nu} K_{b,}^\nu \tilde{T}_a \tilde{T}_b \left(\partial^\mu\Phi\right); \label{art4102h} \\
\La_9 &= \sum_{c=1}^{4} \left(\partial_\mu \Phi\right)^\dagger \ K^\mu_{c,} \tilde{T}_c \ (\Box\Phi); \label{art4102i} \\
\La_{10} &= \sum_{a,c=1}^{4} \left(\partial_\mu \Phi\right)^\dagger \ K^\mu_{c,} (\partial_\nu K_{a,}^\nu) \tilde{T}_c\tilde{T}_a \ \Phi; \label{art4102j} \\
\La_{11} &= 2 \sum_{a,c=1}^{4} \left(\partial_\mu \Phi\right)^\dagger \ K^\mu_{c,} K_{a,}^\nu \tilde{T}_c\tilde{T}_a \ \left(\partial_\nu\Phi\right); \label{art4102k}
\end{align}

\begin{align}
\La_{12} &= \sum_{a,b,c=1}^{4} \left(\partial_\mu \Phi\right)^\dagger \ K_{c,}^\mu  K_{a,\nu} K_{b,}^\nu \tilde{T}_c\tilde{T}_a\tilde{T}_b \ \Phi; \label{art4102l}\\
\La_{13} &= \sum_{d=1}^{4} \Phi^\dagger \ K_{d,\mu}^\dagger \tilde{T}_d^\dagger \ \left(\partial^\mu\Box\Phi\right); \label{art410ma}\\
\La_{14} &= \sum_{a,d=1}^{4} \Phi^\dagger \ K_{d,\mu}^\dagger \left(\partial^\mu\partial_\nu K_{a,}^\nu\right) \tilde{T}_d^\dagger\tilde{T}_a \ \Phi; \label{art4102n}\\
\La_{15} &= \sum_{a,d=1}^{4} \Phi^\dagger \ K_{d,\mu}^\dagger \left(\partial_\nu K_{a,}^\nu\right) \tilde{T}_d^\dagger \tilde{T}_a \ \left(\partial^\mu\Phi\right);\label{art4102o}\\
\La_{16} &= 2 \sum_{a,d=1}^{4} \Phi^\dagger \ K_{d,\mu}^\dagger \left(\partial^\mu\ K_{a,}^\nu\right) \tilde{T}_d^\dagger \tilde{T}_a \ \left(\partial_\nu\Phi\right); \label{art4102p}\\
\La_{17} &= \sum_{a,d=1}^{4} \Phi^\dagger \ K_{d,\mu}^\dagger K_{a,}^\nu \tilde{T}_d^\dagger \tilde{T}_a \ \left(\partial^\mu\partial_\nu\Phi\right);\label{art4102q} \\
\La_{18} &= \sum_{a,b,d=1}^{4} \Phi^\dagger \ K_{d,\mu}^\dagger \left(\partial^\mu K_{a,\nu} \right) K_{b,}^\nu \tilde{T}_d^\dagger \tilde{T}_a \tilde{T}_b \ \Phi; \label{art4102r} \\
\La_{19} &= \sum_{a,b,d=1}^{4} \Phi^\dagger \ K_{d,\mu}^\dagger K_{a,\nu} \left(\partial^\mu K_{b,}^\nu\right) \tilde{T}_d^\dagger \tilde{T}_a \tilde{T}_b \ \Phi; \label{art4102s} \\
\La_{20} &= \sum_{a,b,d=1}^{4} \Phi^\dagger \  K_{d,\mu}^\dagger  K_{a,\nu} K_{b,}^\nu \tilde{T}_d^\dagger \tilde{T}_a \tilde{T}_b \ \left(\partial^\mu\Phi\right); \label{art4102t} \\
\La_{21} &= \sum_{c,d=1}^{4} \Phi^\dagger \ K_{d,\mu}^\dagger  K_{c,}^\mu \tilde{T}_d^\dagger\tilde{T}_c \ \left(\Box \Phi\right); \label{art4102u} \\
\La_{22} &= \sum_{a,c,d=1}^{4} \Phi^\dagger \ K_{d,\mu}^\dagger K_{c,}^\mu (\partial_\nu K_{a,}^\nu) \tilde{T}_d^\dagger \tilde{T}_c\tilde{T}_a \ \Phi; \label{art4102v} \\
\La_{23} &= 2 \sum_{a,c,d=1}^{4} \Phi^\dagger \ K_{d,\mu}^\dagger K_{c,}^\mu K_{a,}^\nu \tilde{T}_d^\dagger \tilde{T}_a \tilde{T}_b \ \left(\partial_\nu\Phi\right); \label{art4102w} \\
\La_{24} &= \sum_{a,b,c,d=1}^{4} \Phi^\dagger \ K_{d,\mu}^\dagger K_{c,}^\mu K_{a,\nu} K_{b,}^\nu\Phi^\dagger \tilde{T}_d^\dagger \tilde{T}_c \tilde{T}_a \tilde{T}_b \ \Phi. \label{art4102x}
\end{align}
As seen in the previous equations (from $\La_1$ to $\La_{24}$), the different contributions to the interaction Lagrangian considered here are given by a product of a factor depending on the electroweak gauge bosons and another one which involves only the Higgs doublet.\\
Basically, each term has the form
\begin{equation}
\label{art4114}
\La_I=F_I \left( K_{a, \nu} \ \text{and \ its \ derivatives} \right) . \ G_I \left( \Phi \ \text{and \ its \ derivatives} \right).
\end{equation}
These functions $G_I$ looking at the formula presented above in this paper have the following forms: 
$$\partial_{\mu} \Phi^\dagger \ \tilde{T} \ \Phi, \ \partial_{\mu} \Phi^\dagger \ \tilde{T} \ \partial^{\mu}\Phi, \ \Phi^\dagger \ \tilde{T} \ \Phi, ...$$
where the matrices $\tilde{T}$ are products of the fundamental matrices $\tilde{T}_1, \tilde{T}_2, \tilde{T}_3, \tilde{T}_4$ (and their hermitian conjugates). Basically, the functions $G_I$ are bilinears in the Higgs doublet and its derivatives.\\

After spontaneous symmetry breaking, one has 
\begin{equation}
\label{art4115}
\Phi=\dfrac{1}{\sqrt{2}} \begin{pmatrix}
0\\
v+h
\end{pmatrix}
=\dfrac{1}{\sqrt{2}} \left(v+h\right) \begin{pmatrix}
0\\
1
\end{pmatrix}.
\end{equation}
The parameter $v$ is knowm experimetally while $h$ is the real scalar field recently discovered as the Higgs boson at the $LHC$. From this one gets,
\begin{equation}
\label{art4116}
\Phi^\dagger=\dfrac{1}{\sqrt{2}} \left(v+h\right) \left(0, 1\right), \ \partial_{\mu}\Phi \dfrac{1}{\sqrt{2}} \left(\partial_{\mu} h\right) \begin{pmatrix}
0\\1
\end{pmatrix}, ...
\end{equation}
With this in mind, let us look at the component of the interaction Lagrangian, we called $\La_2$:
\begin{equation}
\label{art4117}
\La_2 = \sum_{a=1}^{4} \left( \partial^{\mu} \partial_{\nu} K_a, ^\nu \right) \ \partial_{\mu}\Phi^\dagger \ \tilde{T}_a \ \Phi.
\end{equation}
The Higgs part can be rewritten as follows, after spontaneous symmetry breaking:
\begin{eqnarray}
\label{art4118}
\begin{split}
\partial_{\mu} \Phi^\dagger \ \tilde{T}_a \ \Phi &= \dfrac{1}{\sqrt{2}} \left(\partial_{\mu} h\right) (0, 1)
\ \tilde{T}_a \ \dfrac{1}{\sqrt{2}} \left(v+h\right) 
\begin{pmatrix}
0\\
1
\end{pmatrix}\\
&= \dfrac{1}{2} \left(\partial_{\mu} h\right) \left(v+h\right) (0, 1) \ \tilde{T}_a \ \begin{pmatrix}
0\\
1
\end{pmatrix}.
\end{split}
\end{eqnarray}
This naturally leads to the computation of the matrix element $(0, 1) \ \tilde{T}_a \ \begin{pmatrix}
0\\
1
\end{pmatrix}$. $\tilde{T}_a$ is a $2 \times 2$ matrix. It can thus be expressed in terms of its components:
\begin{equation}
\label{art4119}
\tilde{T}_a =\begin{pmatrix}
T_{a, 11} & T_{a, 12}\\
T_{a, 21} & T_{a, 22}
\end{pmatrix},
\end{equation}
one thus has
\begin{eqnarray}
\label{art4120}
\begin{split}
(0, 1) \ \tilde{T}_a \ \begin{pmatrix}
0\\
1
\end{pmatrix} & = (0,1) \begin{pmatrix}
\tilde{T}_{a, 11} & \tilde{T}_{a, 12}\\
\tilde{T}_{a, 21} & \tilde{T}_{a, 22}
\end{pmatrix} 
\begin{pmatrix}
0\\
1
\end{pmatrix}\\
&=(0,1) \begin{pmatrix}
\tilde{T}_{a, 12}\\
\tilde{T}_{a, 22}
\end{pmatrix}\\
&=T_{a, 22}.
\end{split}
\end{eqnarray}
Going back, we find
\begin{equation}
\label{art4121}
\partial_{\mu} \Phi^\dagger \ \tilde{T}_a \ \Phi = \dfrac{1}{2} \left(\partial_{\mu} h\right) \left(v+h\right) \tilde{T}_{a, 22}.
\end{equation}
Introducing the family of matrices $M_1$ such that $M_1(a)=\tilde{T}_a$, one finds with a Mathematica program that
\begin{equation}
\label{art4122}
M_1[a][[2,2]]= i g \sec\theta_W \ \delta_{a, 3}
\end{equation}
one is then led to the following result
\begin{eqnarray}
\label{art4123}
\begin{split}
\La_2 &= \sum_{a=1}^{4} \left( \partial^{\mu} \partial_{\nu} K_a, ^\nu \right) \ \dfrac{1}{2} \partial_{\mu}\Phi^\dagger \left(v+h\right) \ \dfrac{ig}{2} \sec\theta_W \ \delta_{a, 3}\\
&= \dfrac{i}{4} g \sec\theta_W \left( \partial^{\mu} \partial_{\nu} K_3, ^\nu \right) \ \left(\partial_{\mu} h\right) \left(v+h\right)\\
& K_1=W^+, K_2=W^-, K_3=Z, K_4=A\\
\La_2 &=\dfrac{i}{4} g \sec\theta_W \left( \partial^{\mu} \partial_{\nu} Z^\nu \right) \ \left(\partial_{\mu} h\right) \left(v+h\right).
\end{split}
\end{eqnarray}

%%%%%%%%%%%%%%%%%%%%%%%%%%%%%%%%%%%%%%%%%%%%%%%%%%%%%%%%%%%%%%%%%%%%%%%%%%%%%%%%%%%%%%%%%%%%%%%%%%%%%%
%%%%%%%%%%%%%%%%%%%%%%%%%%%%%%%%%%%%%%%%%%%%%%%%%%%%%%%%%%%%%%%%%%%%%%%%%%%%%%%%%%%%%%%%%%%%%%%%%%%%%%

\section{The Matrix Elements}
\label{AppB}

\subsection{Matrix elements involving a single matrix $M_1(a)=\tilde{T}_a$}
\label{AppB01}

The only matrix element is:
\begin{equation}
\label{art4124}
M_1(3)=\dfrac{ig}{2 \cos\theta_W}.
\end{equation}

This element is present in the following terms of the Lagrangian: $\La_2, \La_3, \La_4, \La_5, \La_9$.

%%%%%%%%%%%%%%%%%%%%%%%%%%%%%%%%%%%%%%%%%%%%%%%%%%%%%%%%%%%%%%%%%%%%%%%%%%%%%%%%%%%%%%%%%%%%%%%%%%%%%%%%%%%%%%%%%%%

\subsection{Matrix elements involving a single matrix $\tilde{M}_1(a)=\tilde{T}^\dagger_a$}
\label{AppB02}

The only matrix element is:
\begin{equation}
\label{art4125}
\tilde{M}_1(3)=\dfrac{-ig}{2 \cos\theta_W}
\end{equation}
and it appears in $\La_{13}$.

%%%%%%%%%%%%%%%%%%%%%%%%%%%%%%%%%%%%%%%%%%%%%%%%%%%%%%%%%%%%%%%%%%%%%%%%%%%%%%%%%%%%%%%%%%%%%%%%%%%%%%%%%%%%%%%%%%%

\subsection{Matrix elements involving a product of two matrices $M_2(a,b)=\tilde{T}_a \tilde{T}_b$}
\label{AppB03}

The matrix element are:
\begin{align}
M_2(2, 1)&=\dfrac{-g^2}{2}; \label{art4126a}\\
M_2(3, 3)&=\dfrac{-g^2}{4 \cos\theta_W} \label{art4126b}
\end{align}
and they appear in $\La_6, \La_7, \La_8, \La_{10}, \La_{11}$.

%%%%%%%%%%%%%%%%%%%%%%%%%%%%%%%%%%%%%%%%%%%%%%%%%%%%%%%%%%%%%%%%%%%%%%%%%%%%%%%%%%%%%%%%%%%%%%%%%%%%%%%%%%%%%%%%%%%

\subsection{Matrix elements involving a product of two matrices $\tilde{M}_2(a,b)=\tilde{T}^\dagger_a \tilde{T}_b$}
\label{AppB04}

We found these matrix element:
\begin{align}
\tilde{M}_2(1, 1)&=\dfrac{g^2}{2}; \label{art4127a}\\
\tilde{M}_2(3, 3)&=\dfrac{g^2}{4 \cos\theta_W} \label{art4127b}
\end{align}
and they appear in $\La_{14}, \La_{15}, \La_{16}, \La_{17}, \La_{21}$.

%%%%%%%%%%%%%%%%%%%%%%%%%%%%%%%%%%%%%%%%%%%%%%%%%%%%%%%%%%%%%%%%%%%%%%%%%%%%%%%%%%%%%%%%%%%%%%%%%%%%%%%%%%%%%%%%%%%

\subsection{Matrix elements involving a product of three matrices $M_3(a,b,c)=\tilde{T}_a \tilde{T}_b \tilde{T}_c$}
\label{AppB05}

The matrix element are:
\begin{align}
M_3(2, 1, 3) & = \dfrac{-i g^3}{4 \cos\theta_W}; \label{art4128a}\\
M_3(2, 3, 1) & = \dfrac{i g^3 \cos 2\theta_W}{4\cos\theta_W}; \label{art4128b}\\
M_3(2, 4, 1) & = \dfrac{i e g^2}{2}; \label{art4128c}\\
M_3(3, 2, 1) & = \dfrac{-i g^3}{4 \cos\theta_W}; \label{art4128d}\\
M_3(3, 3, 3) & = \dfrac{-i g^3}{8 \cos^3\theta_W}. \label{art4128e}
\end{align}
They are present in $\La_{12}$.

%%%%%%%%%%%%%%%%%%%%%%%%%%%%%%%%%%%%%%%%%%%%%%%%%%%%%%%%%%%%%%%%%%%%%%%%%%%%%%%%%%%%%%%%%%%%%%%%%%%%%%%%%%%%%%%%%%%

\subsection{Matrix elements involving a product of three matrices $\tilde{M}_3(a,b,c)=\tilde{T}^\dagger_a \tilde{T}_b \tilde{T}_c$}

The matrix element are:
\begin{align}
\tilde{M}_3(1, 1, 3) & = \dfrac{i g^3}{4 \cos\theta_W}; \label{art4129a}\\
\tilde{M}_3(1, 3, 1) & = \dfrac{-i g^3 \cos 2\theta_W}{4\cos\theta_W}; \label{art4129b}\\
\tilde{M}_3(2, 4, 1) & = \dfrac{-i e g^2}{2}; \label{art4129c}\\
\tilde{M}_3(3, 2, 1) & = \dfrac{i g^3}{4 \cos\theta_W}; \label{art4129d}\\
\tilde{M}_3(3, 3, 3) & = \dfrac{i g^3}{8 \cos^3\theta_W}. \label{art4129e}
\end{align}
They are present in these terms of the Lagrangian: $\La_{18}, \La_{19}, \La_{20}, \La_{22}, \La_{23}$.

%%%%%%%%%%%%%%%%%%%%%%%%%%%%%%%%%%%%%%%%%%%%%%%%%%%%%%%%%%%%%%%%%%%%%%%%%%%%%%%%%%%%%%%%%%%%%%%%%%%%%%%%%%%%%%%%%%%

\subsection{Matrix elements involving a product of four matrices $\tilde{M}_4(a,b,c,d)=\tilde{T}^\dagger_a \tilde{T}_b \tilde{T}_c \tilde{T}_d$}

The matrix element are:
\begin{align}
\tilde{M}_4(2, 4, 1, 1) & = \dfrac{-g^4}{4}; \label{art4130a}\\
\tilde{M}_4(3, 1, 3, 1) & = \dfrac{-g^4}{8\cos^2\theta_W}; \label{art4130b}\\
\tilde{M}_4(3, 3, 3, 3) & = \dfrac{-g^4}{8\cos^2\theta_W}; \label{art4130c}\\
\tilde{M}_4(3, 4, 3, 3) & = \dfrac{g^4 \cos 2\theta_W}{8 \cos^2\theta_W}; \label{art4130d}\\
\tilde{M}_4(4, 1, 1, 3) &= \dfrac{e g^3}{4 \cos\theta_W}. \label{art4130e}\\
\tilde{M}_4(4, 2, 3, 1) & = \dfrac{g^4 \cos 2\theta_W}{8 \cos^2\theta_W}; \label{art4130f}\\
\tilde{M}_4(3, 4, 1, 3) & = \dfrac{-g^4 }{8 \cos^2\theta_W}; \label{art4130g}\\
\tilde{M}_4(4, 4, 1, 1) & = \dfrac{-g^4 \cos^2 2\theta_W}{8 \cos^2\theta_W}; \label{art4130h}\\
\tilde{M}_4(4, 4, 3, 3) & = \dfrac{-g^4}{16 \cos^4\theta_W}. \label{art4130i}\\
\end{align}
They are present in this term of the Lagrangian: $\La_{24}$.

%%%%%%%%%%%%%%%%%%%%%%%%%%%%%%%%%%%%%%%%%%%%%%%%%%%%%%%%%%%%%%%%%%%%%%%%%%%%%%%%%%%%%%%%%%%%%%%%%%%%%%
%%%%%%%%%%%%%%%%%%%%%%%%%%%%%%%%%%%%%%%%%%%%%%%%%%%%%%%%%%%%%%%%%%%%%%%%%%%%%%%%%%%%%%%%%%%%%%%%%%%%%%

\section{Noether Theorem}
\label{AppC}

\begin{eqnarray}
\label{art4153}
\left\lbrace\begin{array}{ll}
x^{\mu}\longrightarrow x'^{\mu}=x^\mu+\delta x^\mu\\
\varphi_r(x)\longrightarrow\varphi'_r(x')=\varphi_r(x)+\delta\varphi_r(x)
\end{array}\right. .
\end{eqnarray}
With
\begin{equation}
\label{art4154}
\delta\varphi_r=\varphi'_r(x')-\varphi_r(x).
\end{equation}

\begin{equation}
\label{art415}
\delta\varphi_r(x)=\tilde{\delta}\varphi_r(x)+\frac{\partial\varphi_r(x)}{\partial x^\mu}\delta x^\mu.
\end{equation}

\begin{equation}
\label{art4156}
\partial_\mu\delta\varphi_r(x)=\frac{\partial\varphi_r(x)}{\partial x^\nu}\frac{\partial\delta x^\nu}{\partial x^\mu}+\delta\left(\partial_\mu\varphi_r(x)\right).
\end{equation}

\begin{eqnarray}
\label{art4157}
\begin{split}
\delta S=0&=\int d^4x\left[\frac{\partial\La}{\partial\varphi_r(x)}\tilde{\delta}\varphi_r(x)+\frac{\partial\La}{\partial\partial_\mu\varphi_r(x)}\tilde{\delta}\partial_\mu\varphi_r(x)+\frac{\partial\La}{\partial\left(\partial_\mu\partial_\nu\varphi_r(x)\right)}\tilde{\delta}\left(\partial_\mu\partial_\nu\varphi_r(x)\right)\right.\\
&\left.+\frac{\partial\La}{\partial\left(\partial_\mu\partial_\nu\partial_\sigma\varphi_r(x)\right)}\tilde{\delta}\left(\partial_\mu\partial_\nu\partial_\sigma\varphi_r(x)\right)+\partial_\mu\La\delta x^\mu +\La\partial_\mu\delta x^\mu\right].
\end{split}
\end{eqnarray}
Now, let us explicitly compute the variations of the derivatives of the fields (following the order of appearance) in this last expression; the objective being to have divergences $\partial_\mu\left(...\right)$ and terms in $\tilde{\delta}\varphi_r$.\  
To simplify the writing of the following calculations, let us posit: 
\begin{eqnarray}
\label{art4158}
\begin{split}
A^{r,\mu\nu}&=\frac{\partial\La}{\partial\left(\partial_\mu\partial_\nu\varphi_r(x)\right)} \ ;\\
C^{r,\mu\nu\sigma}&=\frac{\partial\La}{\partial\left(\partial_\mu\partial_\nu\partial_\sigma\varphi_r(x)\right)}.
\end{split}
\end{eqnarray}

\begin{eqnarray}
\label{art4159}
\begin{split}
\frac{\partial\La}{\partial\left(\partial_\mu\varphi_r\right)}\tilde{\delta}\partial_\mu\varphi_r = & \partial_\mu\left(\frac{\partial\La}{\partial\left(\partial_\mu\varphi\right)}\tilde{\delta}\varphi\right)- \partial_\mu\left(\frac{\partial\La}{\partial\left(\partial_\mu\varphi\right)}\right)\tilde{\delta}\varphi_r.
\end{split}
\end{eqnarray} 

\begin{eqnarray*}
	\label{art4160}
	A^{r,\mu\
		nu}\tilde{\delta}\left(\partial_\mu\partial_\nu\varphi_r\right)=A^{r,\mu\nu}\partial_\mu\partial_\nu\left(\tilde{\delta}\varphi_r\right)
\end{eqnarray*}

\begin{eqnarray}
\label{art4161}
\begin{split}
\partial_\mu\partial_\nu\left(A^{r,\mu\nu}\tilde{\delta}\varphi_r\right)&=\left(\partial_\mu\partial_\nu A^{r,\mu\nu}\right)\tilde{\delta}\varphi_r +\left(\partial_\nu A^{r,\mu\nu}\right)\partial_\mu\tilde{\delta}\varphi_r + \left(\partial_\mu A^{r,\mu\nu}\right)\left(\partial_\nu\tilde{\delta}\varphi_r\right)+ A^{r,\mu\nu}\left(\partial_\mu\partial_\nu\tilde{\delta}\varphi_r\right)\\
&=\partial_\mu\left[\partial_\nu\left( A^{r,\mu\nu}+A^{r,\nu\mu}\right)\tilde{\delta}\varphi_r \right]-\partial_\mu\partial_\nu A^{r,\nu\mu}\tilde{\delta}\varphi_r + A^{r,\mu\nu}\left(\partial_\mu\partial_\nu\tilde{\delta}\varphi_r\right).
\end{split}
\end{eqnarray}

\begin{equation}
\label{art4162}
A^{r,\mu\nu}\left(\partial_\mu\partial_\nu\tilde{\delta}\varphi_r\right)=\partial_\mu\left[\partial_\nu\left(A^{r,\mu\nu}\tilde{\delta}\varphi_r\right)-\partial_\nu\left( A^{r,\mu\nu}+A^{r,\nu\mu}\right)\tilde{\delta}\varphi_r \right]+\partial_\mu\partial_\nu A^{r,\nu\mu}\tilde{\delta}\varphi_r.
\end{equation}

\begin{eqnarray*}
	\label{art4163}
	C^{r,\mu\nu\tau}\tilde{\delta}\left(\partial_\mu\partial_\nu\partial_\tau\varphi_r\right)=C^{r,\mu\nu\tau}\partial_\mu\partial_\nu\partial_\tau\left(\tilde{\delta}\varphi_r\right).
\end{eqnarray*}

Using the result from the previous point, we calculate:
\begin{eqnarray}
\label{art4164}
\begin{split}
\partial_\mu\partial_\nu\partial_\tau\left(C^{r,\mu\nu\tau}\tilde{\delta}\varphi_r\right)&=\partial_\mu\partial_\nu\partial_\tau\left( C^{r,\mu\nu\tau}\right)\tilde{\delta}\varphi_r +\partial_\nu\partial_\tau\left( C^{r,\mu\nu\tau}\right)\partial_\mu\tilde{\delta}\varphi_r \\
& +\partial_\mu\partial_\tau\left( C^{r,\mu\nu\tau}+C^{r,\mu\tau\nu}\right)\partial_\nu\tilde{\delta}\varphi_r +\partial_\tau\left( C^{r,\mu\nu\tau}+C^{r,\mu\tau\nu}\right)\partial_\mu\partial_\nu\tilde{\delta}\varphi_r \\
& + \partial_\mu C^{r,\mu\nu\tau}\left(\partial_\nu\partial_\tau\tilde{\delta}\varphi_r\right)+ C^{r,\mu\nu\tau}\left(\partial_\mu\partial_\nu\partial_\tau\tilde{\delta}\varphi_r\right)\\
&=\partial_\mu\partial_\nu\partial_\tau\left( C^{r,\mu\nu\tau}\right)\tilde{\delta}\varphi_r +\partial_\nu\partial_\tau\left( C^{r,\mu\nu\tau}+C^{r,\nu\mu\tau}+C^{r,\nu\tau\mu}\right)\partial_\mu\tilde{\delta}\varphi_r \\
&+\partial_\tau\left( C^{r,\mu\nu\tau}+C^{r,\mu\tau\nu}+C^{r,\tau\nu\mu}\right)\partial_\mu\partial_\nu\tilde{\delta}\varphi_r + C^{r,\mu\nu\tau}\left(\partial_\mu\partial_\nu\partial_\tau\tilde{\delta}\varphi_r\right).
\end{split}
\end{eqnarray}
Let's set
\begin{align}
X^{r,\mu\nu\tau}&=C^{r,\mu\nu\tau}+C^{r,\nu\mu\tau}+C^{r,\nu\tau\mu}\ ; \label{art4165a}\\
Y^{r,\mu\nu\tau}&=C^{r,\mu\nu\tau}+C^{r,\mu\tau\nu}+C^{r,\tau\nu\mu}. \label{art4165b}
\end{align}

\begin{eqnarray}
\label{art4166}
\begin{split}
\partial_\mu\partial_\nu\partial_\tau\left(C^{r,\mu\nu\tau}\tilde{\delta}\varphi_r\right)&=\partial_\mu\partial_\nu\partial_\tau\left[ C^{r,\mu\nu\tau}-X^{r,\mu\nu\tau}\right]\tilde{\delta}\varphi_r +\partial_\mu\left[\partial_\nu\partial_\tau X^{r,\mu\nu\tau}\tilde{\delta}\varphi_r +\partial_\tau Y^{r,\mu\nu\tau}\partial_\nu\tilde{\delta}\varphi_r\right]\\
&-\partial_\nu\left(\partial_\mu\partial_\tau Y^{r,\mu\nu\tau}\tilde{\delta}\varphi_r \right)+\left(\partial_\nu\partial_\mu\partial_\tau Y^{r,\mu\nu\tau}\right)\tilde{\delta}\varphi_r + C^{r,\mu\nu\tau}\left(\partial_\mu\partial_\nu\partial_\tau\tilde{\delta}\varphi_r\right),
\end{split}
\end{eqnarray}

\begin{eqnarray}
\label{art4167}
\begin{split}
\partial_\mu\partial_\nu\partial_\tau\left(C^{r,\mu\nu\tau}\tilde{\delta}\varphi_r\right)&=\partial_\mu\partial_\nu\partial_\tau\left[ C^{r,\mu\nu\tau}-X^{r,\mu\nu\tau}+Y^{r,\mu\nu\tau}\right]\tilde{\delta}\varphi_r \\
&+\partial_\mu\left[\partial_\nu\partial_\tau \left(X^{r,\mu\nu\tau}-Y^{r,\nu\mu\tau}\right)\tilde{\delta}\varphi_r +\partial_\tau Y^{r,\mu\nu\tau}\partial_\nu\tilde{\delta}\varphi_r\right]+ C^{r,\mu\nu\tau}\left(\partial_\mu\partial_\nu\partial_\tau\tilde{\delta}\varphi_r\right).
\end{split}
\end{eqnarray}

\begin{eqnarray}
\label{art4168}
\begin{split}
C^{r,\mu\nu\tau}\left(\partial_\mu\partial_\nu\partial_\tau\tilde{\delta}\varphi_r\right)&= \partial_\mu\partial_\nu\partial_\tau\left[ X^{r,\mu\nu\tau}-Y^{r,\mu\nu\tau}-C^{r,\mu\nu\tau}\right]\tilde{\delta}\varphi_r \\
&+\partial_\mu\left[\partial_\nu\partial_\tau\left(C^{r,\mu\nu\tau}\tilde{\delta}\varphi_r\right)-\partial_\nu\partial_\tau \left(X^{r,\mu\nu\tau}-Y^{r,\nu\mu\tau}\right)\tilde{\delta}\varphi_r -\partial_\tau Y^{r,\mu\nu\tau}\partial_\nu\tilde{\delta}\varphi_r\right].
\end{split}
\end{eqnarray}

\begin{eqnarray*}
	\label{art4169}
	\delta S=0&=&\int d^4x\left\lbrace\frac{\partial\La}{\partial\varphi_r}\tilde{\delta}\varphi_r +\partial_\mu\left(\frac{\partial\La}{\partial\left(\partial_\mu\varphi_r\right)}\tilde{\delta}\varphi_r\right)- \partial_\mu\left(\frac{\partial\La}{\partial\left(\partial_\mu\varphi_r\right)}\right)\tilde{\delta}\varphi_r\right.\\
	&&\qquad+\partial_\mu\left[\partial_\nu\left(A^{r,\mu\nu}\tilde{\delta}\varphi_r\right)-\partial_\nu\left( A^{r,\mu\nu}+A^{r,\nu\mu}\right)\tilde{\delta}\varphi_r \right]+\partial_\mu\partial_\nu A^{r,\nu\mu}\tilde{\delta}\varphi_r\\
	&&\qquad-\partial_\mu\partial_\nu\partial_\tau\left[C^{r,\mu\nu\tau} -X^{r,\mu\nu\tau}+Y^{r,\mu\nu\tau}\right]\tilde{\delta}\varphi_r \\
	&&\qquad\left.+\partial_\mu\left[\partial_\nu\partial_\tau\left(C^{r,\mu\nu\tau}\tilde{\delta}\varphi_r\right)-\partial_\nu\partial_\tau \left(X^{r,\mu\nu\tau}-Y^{r,\nu\mu\tau}\right)\tilde{\delta}\varphi_r -\partial_\tau Y^{r,\mu\nu\tau}\partial_\nu\tilde{\delta}\varphi_r \right]+\partial_\mu\left(\La\delta x^\mu\right)\right\rbrace
\end{eqnarray*}
\begin{eqnarray*}
	\qquad&=&\int d^4x\left\lbrace\frac{\partial\La}{\partial\varphi_r}\tilde{\delta}\varphi_r - \partial_\mu\left(\frac{\partial\La}{\partial\left(\partial_\mu\varphi_r\right)}\right)\tilde{\delta}\varphi_r-\partial_\mu\partial_\nu\partial_\tau\left[C^{r,\mu\nu\tau} -X^{r,\mu\nu\tau}+Y^{r,\mu\nu\tau}\right]\tilde{\delta}\varphi_r \right.\\
	&&\qquad+\partial_\mu\partial_\nu A^{r,\nu\mu}\tilde{\delta}\varphi_r+\partial_\mu\left(\frac{\partial\La}{\partial\left(\partial_\mu\varphi_r\right)}\tilde{\delta}\varphi_r\right)+\partial_\mu\left[ A^{r,\nu\mu}\partial_\nu\tilde{\delta}\varphi_r -\partial_\nu A^{r,\nu\mu}\tilde{\delta}\varphi_r  \right]\\
	&&\qquad\left.+\partial_\mu\left[\partial_\nu\partial_\tau\left(C^{r,\mu\nu\tau}\tilde{\delta}\varphi_r\right)-\partial_\nu\partial_\tau \left(X^{r,\mu\nu\tau}-Y^{r,\nu\mu\tau}\right)\tilde{\delta}\varphi_r -\partial_\tau Y^{r,\mu\nu\tau}\partial_\nu\tilde{\delta}\varphi_r \right]+\partial_\mu\left(\La\delta x^\mu\right)\right\rbrace.
\end{eqnarray*}

\begin{eqnarray}
\label{art4170}
\delta S=\int d^4x\left[\Omega^r\tilde{\delta}\varphi_r +\partial_\mu\Sigma^\mu(x)+\partial_\mu\left(\La\delta x^\mu\right)\right].
\end{eqnarray}

\begin{equation}
\label{art4171}
f^\mu= \Sigma^\mu(x)+\La\delta x^\mu.
\end{equation}

\begin{equation}
\label{art4172}
\delta S=\int d^4x\left[\Omega^r\tilde{\delta}\varphi_r +\partial_\mu f^\mu\right].
\end{equation}

\begin{equation}
\label{art4173}
\Omega^r=\frac{\partial\La}{\partial\varphi_r}- \partial_\mu\left(\frac{\partial\La}{\partial\left(\partial_\mu\varphi_r\right)}\right)+\partial_\mu\partial_\nu A^{r,\nu\mu}-\partial_\mu\partial_\nu\partial_\tau\left[C^{r,\mu\nu\tau} -X^{r,\mu\nu\tau}+Y^{r,\mu\nu\tau}\right],
\end{equation}

\begin{eqnarray}
\label{art4174}
Y^{r,\mu\nu\tau}-X^{r,\mu\nu\tau}&=&C^{r,\mu\tau\nu}+C^{r,\tau\nu\mu}-C^{r,\nu\mu\tau}-C^{r,\nu\tau\mu}
\end{eqnarray}

\begin{eqnarray*}
	\label{art4175}
	\partial_\mu\partial_\nu\partial_\tau\left[-X^{r,\mu\nu\tau}+Y^{r,\mu\nu\tau}\right]&=&\partial_\mu\partial_\nu\partial_\tau C^{r,\mu\tau\nu}+\partial_\mu\partial_\nu\partial_\tau C^{r,\tau\nu\mu}-\partial_\mu\partial_\nu\partial_\tau C^{r,\nu\mu\tau}-\partial_\mu\partial_\nu\partial_\tau C^{r,\nu\tau\mu}\\
	&=&0.
\end{eqnarray*}

\begin{eqnarray*}
	\label{art4176}
	\Omega^r=\frac{\partial\La}{\partial\varphi_r}- \partial_\mu\left(\frac{\partial\La}{\partial\left(\partial_\mu\varphi_r\right)}\right)+\partial_\mu\partial_\nu A^{r,\nu\mu}-\partial_\mu\partial_\nu\partial_\tau C^{r,\mu\nu\tau}=0.
\end{eqnarray*} 

\begin{equation}
\label{art4177}
\frac{\partial\La}{\partial\varphi_r}- \partial_\mu\left(\frac{\partial\La}{\partial\left(\partial_\mu\varphi_r\right)}\right)+\partial_\mu\partial_\nu\left(\frac{\partial\La}{\partial\left(\partial_\mu\partial_\nu\varphi_r(x)\right)}\right)-\partial_\mu\partial_\nu\partial_\tau\left(\frac{\partial\La}{\partial\left(\partial_\mu\partial_\nu\partial_\tau\varphi_r(x)\right)}\right)=0.
\end{equation}

\begin{eqnarray}
\label{art4178}
\begin{split}
\Sigma^\mu(x)&=\frac{\partial\La}{\partial\left(\partial_\mu\varphi_r\right)}\tilde{\delta}\varphi_r +A^{r,\nu\mu}\partial_\nu\tilde{\delta}\varphi_r -\partial_\nu A^{r,\nu\mu}\tilde{\delta}\varphi_r+\partial_\nu\partial_\tau\left(C^{r,\mu\nu\tau}\tilde{\delta}\varphi_r\right)\\
&-\partial_\nu\partial_\tau \left(X^{r,\mu\nu\tau}-Y^{r,\nu\mu\tau}\right)\tilde{\delta}\varphi_r -\partial_\tau Y^{r,\mu\nu\tau}\partial_\nu\tilde{\delta}\varphi_r.
\end{split}
\end{eqnarray} 

\begin{eqnarray}
\label{art4179}
\begin{split}
X^{r,\mu\nu\tau}-Y^{r,\nu\mu\tau}&=C^{r,\mu\nu\tau}+C^{r,\nu\mu\tau}+C^{r,\nu\tau\mu}-C^{r,\nu\mu\tau}-C^{r,\nu\tau\mu}-C^{r,\tau\mu\nu}\\
&=C^{r,\mu\nu\tau}-C^{r,\tau\mu\nu}.
\end{split}
\end{eqnarray}

\begin{eqnarray}
\label{art4180}
\begin{split}
\Sigma^\mu(x)&=\frac{\partial\La}{\partial\left(\partial_\mu\varphi_r\right)}\tilde{\delta}\varphi_r +A^{r,\nu\mu}\partial_\nu\tilde{\delta}\varphi_r -\partial_\nu A^{r,\nu\mu}\tilde{\delta}\varphi_r +\partial_\nu C^{r,\mu\nu\tau}\partial_\tau\tilde{\delta}\varphi_r + C^{r,\mu\nu\tau}\partial_\nu\partial_\tau\tilde{\delta}\varphi_r \\
&+\partial_\nu\partial_\tau C^{r,\tau\mu\nu}\tilde{\delta}\varphi_r -\partial_\tau \left(C^{r,\mu\tau\nu} + C^{r,\tau\nu\mu}\right)\partial_\nu\tilde{\delta}\varphi_r\\
&=\frac{\partial\La}{\partial\left(\partial_\mu\varphi_r\right)}\tilde{\delta}\varphi_r +A^{r,\nu\mu}\partial_\nu\tilde{\delta}\varphi_r -\partial_\nu A^{r,\nu\mu}\tilde{\delta}\varphi_r + C^{r,\mu\nu\tau}\partial_\nu\partial_\tau\tilde{\delta}\varphi_r \\
&+\partial_\nu\partial_\tau C^{r,\tau\mu\nu}\tilde{\delta}\varphi_r -\partial_\tau C^{r,\tau\nu\mu}\partial_\nu\tilde{\delta}\varphi_r.
\end{split}
\end{eqnarray}

\begin{eqnarray}
\label{art4181}
\begin{split}
f^\mu(x) &= \left[\frac{\partial\La}{\partial\left(\partial_\mu\varphi_r\right)}-\partial_\nu A^{r,\nu\mu}+\partial_\nu\partial_\tau C^{r,\tau\mu\nu}\right]\tilde{\delta}\varphi_r\\
&+\left(A^{r,\nu\mu}-\partial_\tau C^{r,\tau\nu\mu}\right)\partial_\nu\tilde{\delta}\varphi_r + C^{r,\mu\nu\tau}\partial_\nu\partial_\tau\tilde{\delta}\varphi_r +\La\delta \ x^\mu.
\end{split}
\end{eqnarray}

\begin{eqnarray}
\label{art4182}
P^{r,\mu} &=&\frac{\partial\La}{\partial\left(\partial_\mu\varphi_r\right)}-\partial_\nu A^{r,\nu\mu}+\partial_\nu\partial_\tau C^{r,\tau\mu\nu}\ ;\label{eq46}\\
R^{r,\mu\nu} &=& A^{r,\nu\mu}-\partial_\tau C^{r,\tau\nu\mu}.
\end{eqnarray}
We have :
\begin{equation}
\label{art4183}
f^\mu(x) = P^{r,\mu}\tilde{\delta}\varphi_r +R^{r,\mu\nu}\partial_\nu\tilde{\delta}\varphi_r + C^{r,\mu\nu\tau}\partial_\nu\partial_\tau\tilde{\delta}\varphi_r +\La \ \delta x^\mu.
\end{equation}

\begin{equation}
\label{art4184}
G=\int_V d^3x \ f^0(x).
\end{equation}

\begin{eqnarray}
\label{art4185}
\left\lbrace\begin{array}{ll}
x'^\mu=x^\mu +\epsilon^\mu\\
\varphi'_r(x')=\varphi_r (x)
\end{array}\right..
\end{eqnarray} 

\begin{equation}
\label{art4186}
f^\mu(x) = P^{r,\mu}\left(\delta\varphi_r-\frac{\partial\varphi_r}{\partial x^\alpha}\delta x^\alpha\right)+R^{r,\mu\nu}\partial_\nu\left(\delta\varphi_r-\frac{\partial\varphi_r}{\partial x^\alpha}\delta x^\alpha\right)
+ C^{r,\mu\nu\tau}\partial_\nu\partial_\tau\left(\delta\varphi_r-\frac{\partial\varphi_r}{\partial x^\alpha}\delta x^\alpha\right) +\La\delta x^\mu.
\end{equation}

\begin{equation}
\label{art4187}
f^\mu(x) = -\left[P^{r,\mu}\partial_\alpha\varphi_r + R^{r,\mu\nu} \partial_\nu\partial_\alpha\varphi_r+C^{r,\mu\nu\tau}\partial_\nu\partial_\tau\partial_\alpha\varphi_r-\eta^\mu_\alpha\La \right]\delta x^\alpha.
\end{equation}

\begin{eqnarray}
\label{art4188}
\begin{split}
\partial_\mu f^\mu(x) =0&= -\partial_\mu\left[P^{r,\mu}\partial_\alpha\varphi_r + R^{r,\mu\nu} \partial_\nu\partial_\alpha\varphi_r+C^{r,\mu\nu\tau}\partial_\nu\partial_\tau\partial_\alpha\varphi_r-\eta^\mu_\alpha\La \right]\delta x^\alpha  \\
&=-\partial_\mu \Theta^\mu_\alpha \delta x^\alpha.
\end{split}
\end{eqnarray}

\begin{eqnarray}
\label{art4189}
\Theta^\mu_\alpha =& P^{r,\mu}\partial_\alpha\varphi_r + R^{r,\mu\nu}    \partial_\nu\partial_\alpha\varphi_r+C^{r,\mu\nu\tau}\partial_\nu\partial_\tau\partial_\alpha\varphi_r-\eta^\mu_\alpha\La. 
\end{eqnarray}

\begin{eqnarray}
\label{art4190}
P_\alpha=\int_V d^3x \ \Theta^0_\alpha. 
\end{eqnarray}

\begin{equation}
\label{art4191}
\Lambda^\mu_\nu=\delta^\mu_\nu +\eta^{\mu\rho}\delta\omega_{\rho\nu}.
\end{equation}

\begin{equation}
\label{art461}
E_r^{\sigma\kappa}=\left[\frac{1}{2}\left(I^{\sigma\kappa}\right)^s_r\varphi_s-\partial^\sigma\varphi_r(x)x^\kappa\right]
\end{equation}

\begin{equation}
\label{art462}
\tilde{\delta}\varphi_r(x) = E_r^{\sigma\kappa}\delta \omega_{\sigma\kappa}.
\end{equation}

\begin{equation}
\label{art463}
f^\mu(x) = \left(\pi^{\mu\sigma\kappa}+\La \eta^{\mu\sigma}x^\kappa\right)\delta \omega_{\sigma\kappa}	
\end{equation}

\begin{equation}
\label{art464}
\partial_\nu E_r^{\sigma\kappa} = F^{s,\sigma\kappa}_{r,\nu\alpha}\partial^\alpha\varphi_s-\partial_\nu\partial^\sigma\varphi_rx^\kappa
\end{equation}

\begin{equation}
\label{art465}
F^{s,\sigma\kappa}_{r,\nu\alpha}=\frac{1}{2}\left(I^{\sigma\kappa}\right)^s_r\eta_{\nu\tau}\delta^\tau_\alpha-\delta^\sigma_\alpha\delta_\nu ^\kappa\delta_r ^s
\end{equation}

\begin{equation}
\label{art466}
\partial_\nu\partial_\tau E_r^{\sigma\kappa} = G^{s,\sigma\kappa\beta}_{r,\nu\tau\alpha}\partial_\beta\partial^\alpha\varphi_s-\partial_\nu\partial_\tau\partial^\sigma\varphi_rx^\kappa.
\end{equation}

\begin{equation}
\label{art467}
G^{s,\sigma\kappa\beta}_{r,\nu\tau\alpha}=F^{s,\sigma\kappa}_{r,\tau\alpha}\delta^\beta_\nu-\delta^\beta_\tau\delta_\alpha^\sigma\delta^s_r\delta^\kappa_\nu.
\end{equation}

\begin{eqnarray}
\label{art468}
\begin{split}
\pi^{\mu\sigma\kappa}&=-\left[P^{r,\mu}\partial^\sigma\varphi_r+R^{r,\mu\nu}\partial_\nu\partial^\sigma\varphi_r+C^{r,\mu\nu\tau}\partial_\nu\partial_\tau\partial^\sigma\varphi_r\right]x^\kappa+\frac{1}{2}P^{r,\mu}\left(I^{\sigma\kappa}\right)^s_r\varphi_s\\
&+ R^{r,\mu\nu}\left[\frac{1}{2}\left(I^{\sigma\kappa}\right)^s_r\eta_{\nu\tau}\delta^\tau_\alpha-\delta_r ^s\delta^\sigma_\alpha\delta_\nu ^\kappa\right]\partial^\alpha\varphi_s + C^{r,\mu\nu\tau}\left[F^{s,\sigma\kappa}_{r,\tau\alpha}\delta^\beta_\nu-\delta^\beta_\tau\delta_\alpha^\sigma\delta^s_r\delta^\kappa_\nu\right]\partial_\beta\partial^\alpha\varphi_s\\
&=-\left[\Theta^{\mu\sigma}+\eta^{\mu\sigma}\La\right]  x^\kappa+\frac{1}{2}\left(I^{\sigma\kappa}\right)^s_r\left[P^{r,\mu}\varphi_s+R^{r,\mu\nu}\eta_{\nu\alpha}\partial^\alpha\varphi_s+C^{r,\mu\nu\tau}\eta_{\tau\alpha}\partial_\nu\partial^\alpha\varphi_s\right]\\
&-R^{r,\mu\kappa}\partial^\sigma\varphi_r-C^{r,\mu\kappa\tau} \partial_\tau\partial^\sigma\varphi_r+-C^{r,\mu\nu\kappa} \partial_\nu\partial^\sigma\varphi_r.
\end{split}
\end{eqnarray}

\begin{eqnarray}
\label{art469}
\begin{split}
f^\mu(x)&=\frac{1}{2}\left[P^{r,\mu}\varphi_s+R^{r,\mu\nu}\eta_{\nu\tau}\delta^\tau_\alpha\partial^\alpha\varphi_s+C^{r,\mu\nu\tau}\eta_{\tau\gamma}\delta^\gamma_\alpha\delta^\beta_\nu\partial_\beta\partial^\alpha\varphi_s\right]\left(I^{\sigma\kappa}\right)^s_r\delta \omega_{\sigma\kappa}\\
&+\frac{1}{2}\left[\left(\Theta^{\mu\kappa}x^\sigma-\Theta^{\mu\sigma}x^\kappa\right)-2R^{r,\mu\kappa}\partial^\sigma\varphi_r-2C^{r,\mu\kappa\tau} \partial_\tau\partial^\sigma\varphi_r-2C^{r,\mu\nu\kappa} \partial_\nu\partial^\sigma\varphi_r\right]\delta \omega_{\sigma\kappa}.
\end{split}
\end{eqnarray}

\begin{eqnarray}
\label{art470}
\begin{split}
\Theta^{\mu\sigma}x^\kappa\delta \omega_{\sigma\kappa}&=\frac{1}{2}\left[\Theta^{\mu\sigma}x^\kappa\delta \omega_{\sigma\kappa}+\Theta^{\mu\kappa}x^\sigma\delta \omega_{\kappa\sigma}\right]\\
&=\frac{1}{2}\left[\Theta^{\mu\sigma}x^\kappa-\Theta^{\mu\kappa}x^\sigma\right]\delta \omega_{\sigma\kappa},
\end{split}
\end{eqnarray}
ainsi
\begin{eqnarray}
\label{art471}
\begin{split}
f^\mu(x)&=\frac{1}{2}\left[P^{r,\mu}\varphi_s+R^{r,\mu\nu}\eta_{\nu\tau}\delta^\tau_\alpha\partial^\alpha\varphi_s+C^{r,\mu\nu\tau}\eta_{\tau\gamma}\delta^\gamma_\alpha\delta^\beta_\nu\partial_\beta\partial^\alpha\varphi_s\right]\left(I^{\sigma\kappa}\right)^s_r\delta \omega_{\sigma\kappa}\\
&+\frac{1}{2}\Big[\left(\Theta^{\mu\kappa}x^\sigma-\Theta^{\mu\sigma}x^\kappa\right)-2R^{r,\mu\kappa}\partial^\sigma\varphi_r-2C^{r,\mu\kappa\tau} \partial_\tau\partial^\sigma\varphi_r-2C^{r,\mu\nu\kappa} \partial_\nu\partial^\sigma\varphi_r\Big]\delta \omega_{\sigma\kappa}.
\end{split}
\end{eqnarray}

\begin{equation}
\label{art472}
f^\mu(x) = \frac{1}{2}M^{\mu\sigma\kappa}\delta \omega_{\sigma\kappa}
\end{equation}

\begin{eqnarray}
\label{art473}
\begin{split}
M^{\mu\sigma\kappa}&=\left[P^{r,\mu}\varphi_s+R^{r,\mu\nu}\eta_{\nu\tau}\delta^\tau_\alpha\partial^\alpha\varphi_s+C^{r,\mu\nu\tau}\eta_{\tau\gamma}\delta^\gamma_\alpha\delta^\beta_\nu\partial_\beta\partial^\alpha\varphi_s\right]\left(I^{\sigma\kappa}\right)^s_r\\
&+\Big[\left(\Theta^{\mu\kappa}x^\sigma-\Theta^{\mu\sigma}x^\kappa\right)-2R^{r,\mu\kappa}\partial^\sigma\varphi_r-2C^{r,\mu\kappa\tau} \partial_\tau\partial^\sigma\varphi_r-2C^{r,\mu\nu\kappa} \partial_\nu\partial^\sigma\varphi_r \Big].
\end{split}
\end{eqnarray}

\begin{eqnarray}
\label{art474}
\begin{split}
M^{\sigma\kappa}&=\int d^3x\left[P^{r,\mu}\varphi_s+R^{r,\mu\nu}\eta_{\nu\tau}\delta^\tau_\alpha\partial^\alpha\varphi_s+C^{r,\mu\nu\tau}\eta_{\tau\gamma}\delta^\gamma_\alpha\delta^\beta_\nu\partial_\beta\partial^\alpha\varphi_s\right]\left(I^{\sigma\kappa}\right)^s_r\\
&+\int d^3x \Big[\left(\Theta^{\mu\kappa}x^\sigma-\Theta^{\mu\sigma}x^\kappa\right)-2R^{r,\mu\kappa}\partial^\sigma\varphi_r-2C^{r,\mu\kappa\tau} \partial_\tau\partial^\sigma\varphi_r-2C^{r,\mu\nu\kappa} \partial_\nu\partial^\sigma\varphi_r \Big]\\
&=L^{\sigma\kappa}+S^{\sigma\kappa}
\end{split}
\end{eqnarray}

\begin{equation}
\label{art475}
\varphi'_r(x) = \varphi_r(x)+i\epsilon\lambda_r^s\varphi_s.
\end{equation}

\begin{eqnarray}
\label{art476}
f^\mu(x)=i\epsilon \ \lambda_r^s \Big[P^{r,\mu}\varphi_s+R^{r,\mu\nu}\partial_\nu\varphi_s+C^{r,\mu\nu\tau}\partial_\nu\partial_\tau\varphi_s \Big].
\end{eqnarray}

\begin{equation}
\label{art477}
Q=\int d^3x\lambda_r^s\left[P^{r,0}\varphi_s+R^{r,0\nu}\partial_\nu\varphi_s+C^{r,0\nu\tau}\partial_\nu\partial_\tau\varphi_s\right].
\end{equation}

\begin{align}
L^{\sigma\kappa}&=\int d^3x \Big[\left(\Theta^{0\kappa}x^\sigma-\Theta^{0\sigma}x^\kappa\right)-2R^{r,0\kappa}\partial^\sigma\varphi_r-2\left(C^{r,0\kappa\tau}\partial_\tau\partial^\sigma\varphi_r+C^{r,0\nu\kappa}\partial_\nu\partial^\sigma\varphi_r\right) \Big]; \label{art478a}\\
S^{\sigma\kappa}&=\int d^3x \Big[P^{r,0}\varphi_s+R^{r,0\nu}\partial_\nu\varphi_s+C^{r,0\nu\tau}\partial_\nu\partial_\tau\varphi_s \Big]\left(I^{\sigma\kappa}\right)^s_r.\label{art478b}
\end{align}

%%%%%%%%%%%%%%%%%%%%%%%%%%%%%%%%%%%%%%%%%%%%%%%%%%%%%%%%%%%%%%%%%%%%%%%%%%%%%%%%%%%%%%%%%%%%%%%%%%%%%%%%%%%%%%%%%%%%%%%%%%%%%
%%%%%%%%%%%%%%%%%%%%%%%%%%%%%%%%%%%%%%%%%%%%%%%%%%%%%%%%%%%%%%%%%%%%%%%%%%%%%%%%%%%%%%%%%%%%%%%%%%%%%%%%%%%%%%%%%%%%%%%%%%%%%

\end{document}